\documentclass[twocolumn,tighten,trackchanges]{aastex62}
\usepackage{graphicx}		
\usepackage{url}
\usepackage{epstopdf}
\usepackage{color}
\usepackage{multirow}
\usepackage{ragged2e}
\usepackage{hyperref}
\usepackage{url}
\usepackage{xspace}
\DeclareGraphicsExtensions{.pdf,.png,.jpeg}
\def\sgra{Sgr\,A$^{\ast}$}
\def\lsim{\mathrel{\raise.3ex\hbox{$<$\kern-.75em\lower1ex\hbox{$\sim$}}}}
\def\gsim{\mathrel{\raise.3ex\hbox{$>$\kern-.75em\lower1ex\hbox{$\sim$}}}}
\def\gtwid{\mathrel{\raise.3ex\hbox{$>$\kern-.75em\lower1ex\hbox{$\sim$}}}}
\def\proptwid{\mathrel{\raise.3ex\hbox{$\propto$\kern-.75em\lower1ex\hbox{$\sim$}}}}
\DeclareUnicodeCharacter{2212}{-}
\newcommand{\themis}{{\sc Themis}\xspace} 

\defcitealias{Johnson_2018}{J18}
\defcitealias{Brinkerink_2019}{B19}
\defcitealias{Brinkerink_2016}{B16}
\defcitealias{Ortiz_2016}{O16}
\defcitealias{Goldreich_2006}{GS06}

\def\apjl{ApJL}
\def\apjs{ApJS}
\def\ao{Appl.~Opt.}
\def\aap{A\&A}
\newcommand{\naoj}{National Astronomical Observatory of Japan, 2-21-1 Osawa, Mitaka, Tokyo 181-8588, Japan}
\newcommand{\haystack}{Massachusetts Institute of Technology, Haystack Observatory, 99 Millstone Rd, Westford, MA 01886, USA}
\newcommand{\bhi}{Black Hole Initiative, Harvard University, 20 Garden Street, Cambridge, MA 02138, USA}
\newcommand{\perimeter}{Perimeter Institute for Theoretical Physics, 31 Caroline Street, North Waterloo, Ontario N2L 2Y5, Canada}
\newcommand{\asiaa}{Institute of Astronomy and Astrophysics, Academia Sinica, P.O. Box 23-141, Taipei 10617, Taiwan}
\newcommand{\radboud}{Department of Astrophysics/IMAPP, Radboud University, P.O. Box 9010, 6500 GL Nijmegen, The Netherlands}

\newcommand{\cfa}{Center for Astrophysics $|$ Harvard \& Smithsonian, 60 Garden Street, Cambridge, MA 02138, USA}
\newcommand{\mpifr}{Max-Planck-Institut f{\"u}r Radioastronomie, Auf dem H{\"u}gel 69, D-53121 Bonn, Germany}

\newcommand{\inaoe}{Instituto Nacional de Astrof{\'i}sica {\'O}ptica y Electr{\'o}nica (INAOE), Apartado Postal 51 y 216, 72000, Puebla, M{\'e}xico}

\newcommand{\frankfurt}{Institut f{\"u}r Theoretische Physik, Johann Wolfgang Goethe-Universit{\"a}t, Max-von-Laue-Stra{\ss}e 1, 60438 Frankfurt, Germany}
\newcommand{\arizona}{University of Arizona, 933 North Cherry Avenue, Tucson, AZ 85721, USA}
\newcommand{\granada}{Instituto de Astrof{\'i}sica de Andaluc{\'i}a-CSIC, Glorieta de la Astronom{\'i}a s/n, E-18008 Granada, Spain}
\newcommand{\kasi}{Korea Astronomy and Space Science Institute, Daedeok−daero 776, Yuseong−gu, Daejeon 34055, Korea}
\newcommand{\api}{Anton Pannekoek Institute for Astronomy, University of Amsterdam, 1098 XH Amsterdam, The Netherlands}

\newcommand{\shanghai}{Shanghai Astronomical Observatory, Chinese Academy of Sciences, Shanghai 200030, China}
\newcommand{\unam}{Instituto de Radioastronom{\'i}a y Astrof{\'i}sica, Universidad Nacional Aut{\'o}noma de Mexico, Morelia 58089, M{\'e}xico }
\newcommand{\unamm}{Instituto de Astronom{\'i}a, Universidad Nacional Aut{\'o}noma de M{\'e}xico, Apartado Postal 70-264, 04510 Ciudad de M{\'e}xico, M{\'e}xico }
\newcommand{\ust}{University of Science and Technology, Gajeong-ro 217, Yuseong-gu, Daejeon 34113, Korea}

\newcommand{\bologna}{Italian ALMA Regional Centre, INAF-Istituto di Radioastronomia, Via P. Gobetti 101, I-40129 Bologna, Italy}
\newcommand{\jiatong}{Tsung-Dao Lee Institute, Shanghai Jiao Tong University, Shanghai, 200240, China}
\newcommand{\princeton}{Princeton Center for Theoretical Science, Jadwin Hall, Princeton University, Princeton, NJ 08544, USA}

\begin{document}

\title{Persistent Non-Gaussian Structure in the Image of Sagittarius A* at 86 GHz}
\shorttitle{Persistent Non-Gaussian Structure in Sgr A*}

\author{S.~Issaoun}
\affil{\radboud}
\author{M.~D.~Johnson}
\affil{\cfa}
\author{L.~Blackburn}                               
\affil{\cfa}
\author{A.~Broderick}
\affil{\perimeter}
\author{P.~Tiede}
\affil{\perimeter}
\author{M.~Wielgus}                                 
\affil{\bhi}\affil{\cfa}
 \author{ S.~S.~Doeleman}                       
 \affil{\cfa}
\author{H.~Falcke}
\affil{\radboud}
 \author{K.~Akiyama}                             
\affil{\haystack}
\author{G.~C.~Bower}                            
\affil{\asiaa}
\author{C.~D.~Brinkerink}
\affil{\radboud}
\author{A.~Chael}                               
\affil{\princeton}
\author{I.~Cho}                                 
\affil{\kasi}\affil{\ust}
\author{J.~L.~G{\'o}mez}                        
\affil{\granada}
 \author{A.~Hern{\'a}ndez-G{\'o}mez}            
\affil{\mpifr}
\author{D.~Hughes}                              
\affil{\inaoe}
\author{M.~Kino}                                
\affil{\naoj}\affil{Kogakuin University of Technology \& Engineering, Academic Support Center, 2665-1 Nakano, Hachioji, Tokyo 192-0015, Japan}
\author{T.~P.~Krichbaum}                       
\affil{\mpifr}
\author{E. Liuzzo}                             
\affil{\bologna}
\author{L.~Loinard}                           
\affil{\unam}\affil{\unamm}
\author{S.~Markoff}                           
\affil{\api}
\author{D.~P.~Marrone}                      
\affil{\arizona}
\author{Y. Mizuno}                        
\affil{\frankfurt}\affil{\jiatong}
\author{J.~M.~Moran}                      
\affil{\cfa}
 \author{Y.~Pidopryhora}                
\affil{\mpifr}
\author{E.~Ros}                         
\affil{\mpifr}
 \author{K.~Rygl}                       
 \affil{\bologna}
\author{Z.-Q.~Shen}                      
\affil{\shanghai}
 \author{J.~Wagner}                
 \affil{\mpifr}

\newcommand{\mw}[1]{\textcolor{teal}{MW: #1}}
\newcommand{\ac}[1]{\textcolor{purple}{AC: #1}}
\newcommand{\lb}[1]{\textcolor{blue}{LB: #1}}
\newcommand{\mdj}[1]{\textcolor{olive}{MDJ: #1}}
\newcommand{\si}[1]{\textcolor{magenta}{SI: #1}}
\newcommand{\pt}[1]{\textcolor{cyan}{PT: #1}}
\newcommand{\aeb}[1]{\textcolor{green}{AB: #1}}
\newcommand{\edt}[1]{\textcolor{red}{#1}}
\newcommand{\todo}[1]{\textcolor{orange}{TO DO - #1}}


\begin{abstract}
Observations of the Galactic Center supermassive black hole Sagittarius A* (\sgra) with very long baseline interferometry (VLBI) are affected by interstellar scattering along our line of sight. At long radio observing wavelengths ($\gsim1\,$cm), the scattering heavily dominates image morphology. At 3.5\,mm (86\,GHz), the intrinsic source structure is no longer sub-dominant to scattering, and thus the intrinsic emission from \sgra\ is resolvable with the Global Millimeter VLBI Array (GMVA). Long-baseline detections to the phased Atacama Large Millimeter/submillimeter Array (ALMA) in 2017 provided new constraints on the intrinsic and scattering properties of \sgra, but the stochastic nature of the scattering requires multiple observing epochs to reliably estimate its statistical properties. We present new observations with the GMVA+ALMA, taken in 2018, which confirm non-Gaussian structure in the scattered image seen in 2017. In particular, the ALMA--GBT baseline shows more flux density than expected for an anistropic Gaussian model, providing a tight constraint on the source size and an upper limit on the dissipation scale of interstellar turbulence.
We find an intrinsic source extent along the minor axis of $\sim100\,\mu$as both via extrapolation of longer wavelength scattering constraints and direct modeling of the 3.5\,mm observations. Simultaneously fitting for the scattering parameters, we find an at-most modestly asymmetrical (major-to-minor axis ratio of $1.5\pm 0.2$) intrinsic source morphology for \sgra. 
\end{abstract}


\section{Introduction}
The Galactic Center hosts the closest known supermassive black hole (SMBH), associated with the radio source Sagittarius A* \citep[\sgra;][]{Balick_Brown_1974}. With a mass $M\sim4.1\times10^6M_\odot$ at a distance $D\sim8.1\,$kpc, \sgra\ subtends the largest angle on the sky among all known black holes \citep{Ghez_2008,Reid_2009,Gillessen_2009,Gravity_2018}. Thus \sgra\ is one of the most promising targets to study black hole accretion and outflow via direct imaging \citep{Goddi_2016}. The spectral energy density of \sgra\ in radio rises with frequency, with a turnover in the sub-millimeter regime, where the accretion flow becomes optically thin~\citep{Falcke_1998,Bower_2015,Bower_2019}.  
However, the southern declination and interstellar scattering of \sgra\ add challenges to decades of radio observations with very long baseline interferometry \citep[VLBI;][]{Alberdi_1993,Backer_1993,Krichbaum_1993,Marcaide_1999,Bower_2004,Shen_2005,Lu_2011,Bower_2014b}. Thus, the intrinsic accretion and outflow structure of \sgra\ remains rather poorly understood. 

Early observations at 1.3\,mm with the prototype Event Horizon Telescope (EHT) indicate that the radio emission of \sgra\ originates from a region that is comparable to the size of the black hole's ``shadow''  \citep[$\sim 50\,\mu$as;][]{Doeleman_2008,Fish_2011,Johnson_2015,Fish_2016,Lu_2018}. On these scales, the image morphology is dominated by strong gravitational lensing of the black hole 
rather than by details of the innermost accretion flow \citep[as seen in M87;][]{PaperI}. At longer wavelengths, images of \sgra\ are strongly scatter-broadened (blurred) by the intervening interstellar medium \citep[ISM; e.g.,][]{Davies_1976,vanLangevelde_1992,Frail_1994,Bower_2004,Shen_2005,Bower_2006,Psaltis_2018,Johnson_2018}. 

Radio waves passing through the ionized ISM propagate via multiple paths due to changes in the refractive index of the turbulent plasma from density inhomogeneities. The angles at which the waves scatter are proportional to the squared wavelength of the wave. The intrinsic angular size of \sgra\ at wavelengths of 0.1--1\,centimeters is roughly proportional to the wavelength. 
As a result, the ratio of intrinsic source angular size to scatter-broadening is ${{\sim}0.3}/\lambda_{\rm cm}$ along the major axis and ${{\sim}0.6}/\lambda_{\rm cm}$ along the minor axis \citep[where $\lambda_{\rm cm}$ is the observing wavelength in centimeters;][]{Johnson_2018}, making 3.5\,mm the longest observing wavelength accessible on Earth at which \sgra\ intrinsic structure would not be sub-dominant to scattering.
 The ideal regime to probe and separate intrinsic source properties from scattering is thus at 3.5\,mm: intrinsic structure starts to dominate over scattering effects, and the radio emission originates from the optically thick innermost accretion flow, also corresponding to the launching region of a possible outflow or jet \citep{Narayan_1995,Falcke_Markoff_2000,Ozel_2000,Yuan_2003}.

At 1.3\,mm, the effects of scattering can additionally potentially contaminate tests of general relativity with the EHT, introducing random distortions and substructure in the image. The specific effects on 1.3\,mm VLBI images depend on the power spectrum $\mathcal{Q}(\mathrm{\bf q})$ of spatial irregularities that produce the scattering \citep[where $\mathrm{\bf q}$ is a wavevector;][]{Johnson_2018,Zhu_2018}. Because these underlying irregularities that cause refractive scattering at 3.5\,mm also produce image variations at 1.3\,mm, scattering studies at 3.5\,mm are essential to guide imaging \sgra\ at 1.3\,mm with the EHT.
Furthermore, scattering-induced substructure, predicted by \citet{NarayanGoodman89} and \citet{GoodmanNarayan89} and first measured in \sgra\ by \citet{Gwinn_2014} at 1.3\,cm, is caused by modes in the scattering material on scales much larger than the diffractive scale of the scattering. 
This turbulence in the ISM induces stochastically varying compact substructure in images of \sgra\ that contaminates long-baseline source behavior with added ``refractive noise'' in the visibility domain, making the recovery of small-scale intrinsic source structure difficult \citep{Johnson_Gwinn_2015,Johnson_Narayan_2016}. 

\begin{figure*}[ht]
\centering
\includegraphics[width=0.8\linewidth]{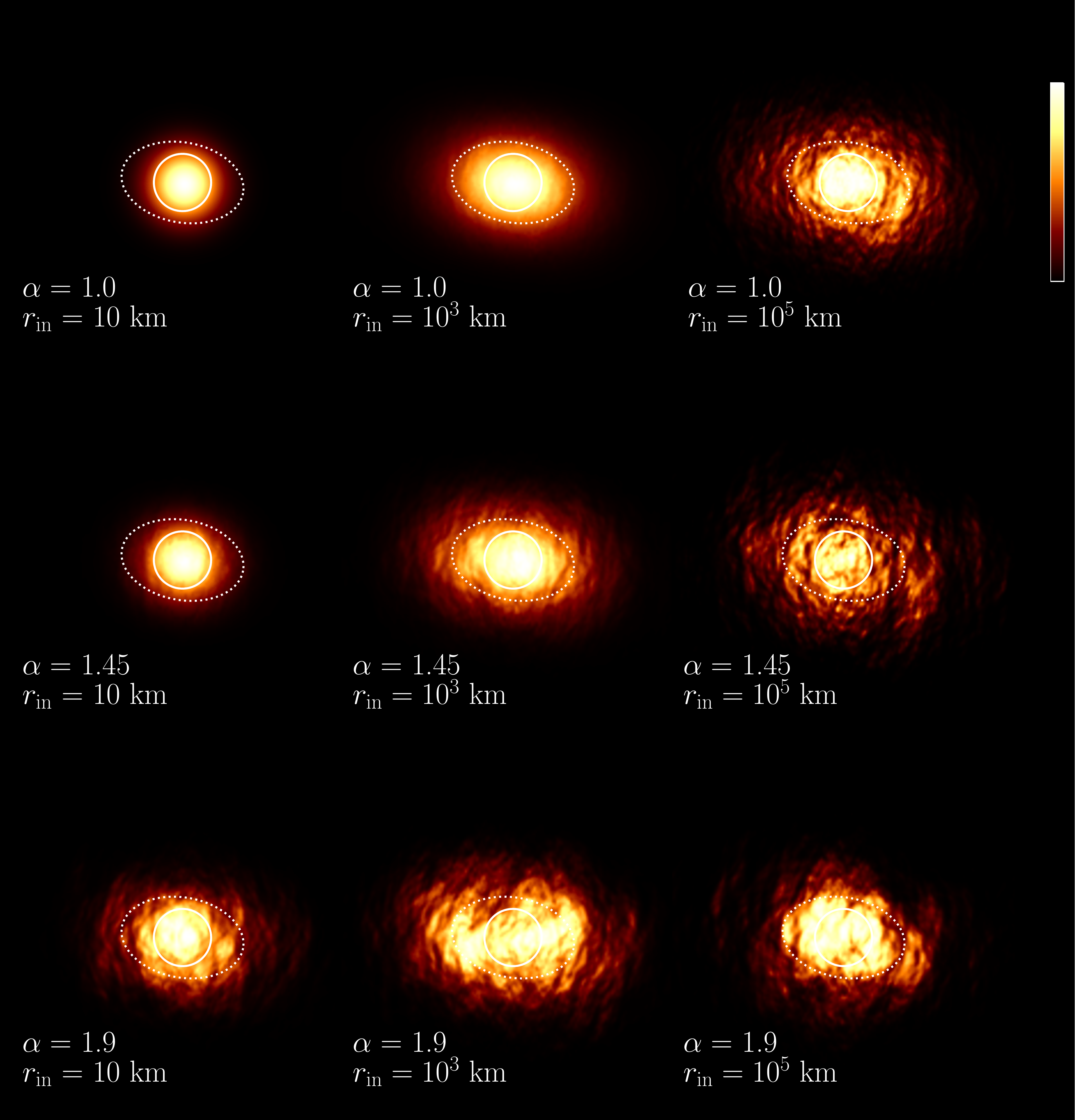}
\caption{Modeled effects of interstellar scattering on a simulated source whose intrinsic structure is that of a circular Gaussian with a full-width at half-maximum (FWHM) of 100\,$\mu$as (shown as the solid circle) at $\lambda=3.5$\,mm with varying power-law index $\alpha$ and inner scale of turbulence $r_\mathrm{in}$. The dashed ellipse shows the FWHM size of the measured Gaussian source on the sky for \sgra\ in previous 3.5\,mm experiments \citep{Ortiz_2016,Brinkerink_2019}. Each row has constant $\alpha$ and varying $r_\mathrm{in}$ (10, $10^3$, and $10^5$\,km): as $r_\mathrm{in}$ increases, the scatter-broadening and the level of refractive substructure increase. Each column has constant $r_\mathrm{in}$ and varying $\alpha$ (1.0, 1.45, and 1.9): as $\alpha$ increases, the scatter-broadening and the level of refractive substructure increase.
}
\label{fig:scatt_grid}
\end{figure*}

In this paper, we utilize the scattering model developed by \citet{Psaltis_2018}, using physical parameters from \citet{Johnson_2018} that were estimated using archival observations of \sgra. The two-dimensional power spectrum of the phase fluctuations $\mathcal{Q}(\mathrm{\bf q})$ is modeled as an unbroken anisotropic power-law with a spectral index $\beta$ extending between a maximum scale (the outer scale $r_{\rm out}$) and a minimum scale (the inner scale $r_{\rm in}$): $Q(\mathrm{\bf q}) \propto |\mathrm{\bf q}|^{-\beta}$ \citep[$\beta$ is also the exponent for the three-dimensional power spectrum of density fluctuations; e.g.,][]{Blandford_Narayan_1985,Rickett_1990}. This power spectrum then yields a second-order phase structure function $D_\phi(\mathbf{r}) = \left \langle \left[ \phi(\mathbf{r}' + \mathbf{r}) - \phi(\mathbf{r}') \right]^2 \right \rangle \propto \left| \mathbf{r} \right|^\alpha$ in the inertial range $r_{\rm in} \ll r \ll r_{\rm out}$, where $\alpha \equiv \beta - 2$. While two scattering models may have identical scatter-broadening, they may still differ wildly in their refractive substructure. Combining information from both scatter-broadening from 1.3\,mm to 30\,cm and centimeter-wave substructure strongly constrains the scattering power spectrum and the asymptotic Gaussian morphology parameters of the scatter-broadening kernel. However, a degeneracy between the power-law index $\alpha$ and the inner scale of the turbulence in the ISM $r_{\rm in}$ remains \citep{Johnson_2018}: various combinations of scattering and intrinsic source parameters can produce the same observed behavior in the scattered image, illustrated in Figure~\ref{fig:scatt_grid}.
Sensitive VLBI observations at 3.5\,mm offer a prime opportunity to break degeneracies between the parameters by connecting to millimeter-wave scattering behavior.

For the past two decades since its first detection at 3.5\,mm \citep{Rogers_1994}, the scattered image of \sgra\ has been commonly modeled as an elliptical Gaussian source with a position angle of ${\sim}\,80^\circ$ east of north \citep[e.g.,][]{Shen_2005,Bower_2006,Lu_2011,Ortiz_2016,Brinkerink_2019}. Closure phases --- the directed sums of visibility phases over closed triangles of baselines
and robust to station-based errors \citep[e.g,][]{Cornwell_1989,Rauch_2016,TMS,lindyclosures} --- measured by early 3.5\,mm experiments were consistently zero \citep{Rogers_1994,Krichbaum_1998,Doeleman_2001,Shen_2005,Bower_2006,Lu_2011}, indicating source symmetry on the probed spatial scales, while non-zero closure phases at lower frequencies are entirely attributable to interstellar scattering \citep{Johnson_2018}. 
A new set of higher-sensitivity experiments, including the Large Millimeter Telescope Alfonso Serrano (LMT) and the Robert C. Byrd Green Bank Telescope (GBT), detected the first non-zero 3.5\,mm closure phases in \sgra\ on new closure triangles provided by the addition of the LMT, which could either be due to intrinsic structure or non-Gaussian structure in the scattering screen \citep{Ortiz_2016,Brinkerink_2016}.
These results motivate breaking the assumptions of an elliptical Gaussian source model and attempting to recover complex underlying source structure via imaging. 

The recently added VLBI phasing capability to the Atacama Large Millimeter/submillimeter Array (ALMA) provided additional sensitivity and long north-south baselines to the Global Millimeter VLBI Array (GMVA) at 3.5\,mm \citep{Matthews_2018}.
While pre-ALMA experiments could not identify the detailed morphology or constrain the radio emission model, the sensitivity and coverage brought by joining ALMA to the GMVA for the first time in 2017 --- including tripling the angular north-south resolution --- has offered a major leap in imaging capabilities and model discrimination for \sgra. 
In \citet{Issaoun_2019}, we showed that measured visibility amplitudes on long baselines to ALMA exhibit clear non-Gaussian behavior, which was a function of baseline length, and ruled out a potential scattering model that would significantly contaminate future EHT images at 1.3\,mm \citep{Zhu_2018}.
Using interstellar scattering mitigation methods \citep{Johnson_2016} coupled with the enhanced coverage of GMVA+ALMA, we then reconstructed a first image of the unscattered structure of \sgra\ at 3.5\,mm. The unscattered source had a major axis full width at half-maximum (FWHM) of $120 \pm 34\,\mu$as ($12.0 \pm 3.4$\,Schwarzschild radii ($R_\mathrm{Sch}$); where $R_\mathrm{Sch} = 2\,GM/c^2$) and a circularly symmetric morphology (major-to-minor-axis ratio of $1.2_{-0.2}^{+0.3}$), which requires either that the accretion flow dominates the emission or that jet-dominant emission from \sgra\ is pointed within $20^\circ$ of the line of sight.

\citet{Issaoun_2019} used the scattering model that best matched observations \citep[][hereafter model J18]{Johnson_2018} in the imaging process to mitigate the scattering effects and recover the intrinsic structure.
Based on a single observation, it is not clear whether the scattering parameter assumptions for \sgra\ are valid: long-baseline detections could either be attributed to intrinsic structure, scattering substructure, or a mix of both. The 2017 long-baseline detections were more consistent with the near-Kolmogorov power spectrum (power-law index $\alpha=1.38$) from \citetalias{Johnson_2018} rather than a flat spectrum \citep[][hereafter model GS06]{Goldreich_2006} governing the stochastic variations in the refractive noise. It is however possible that the low refractive noise observed can be attributed to a statistically unlikely low refractive noise realization from the \citetalias{Goldreich_2006} flat-spectrum scattering model. Thus, it remains important to sample these scales at different instances in time.
We therefore performed follow-up observations of \sgra\ with GMVA+ALMA in 2018 to gain further confidence in the scattering model and tighten constraints in the model parameters to describe the ISM along the line of sight to \sgra.

The organization of the paper is as follows. We summarize observations and data reduction in Section~\ref{sec:data}, present our final GMVA+ALMA visibility amplitudes on \sgra\ in Section~\ref{sec:results}, and discuss the constraints on scattering and intrinsic source parameters enabled by these latest results in Section~\ref{sec:discussion}. A summary is given in Section~\ref{sec:summary}.

\begin{figure}[t]
\hspace{1.5mm}\includegraphics[width=0.97\linewidth]{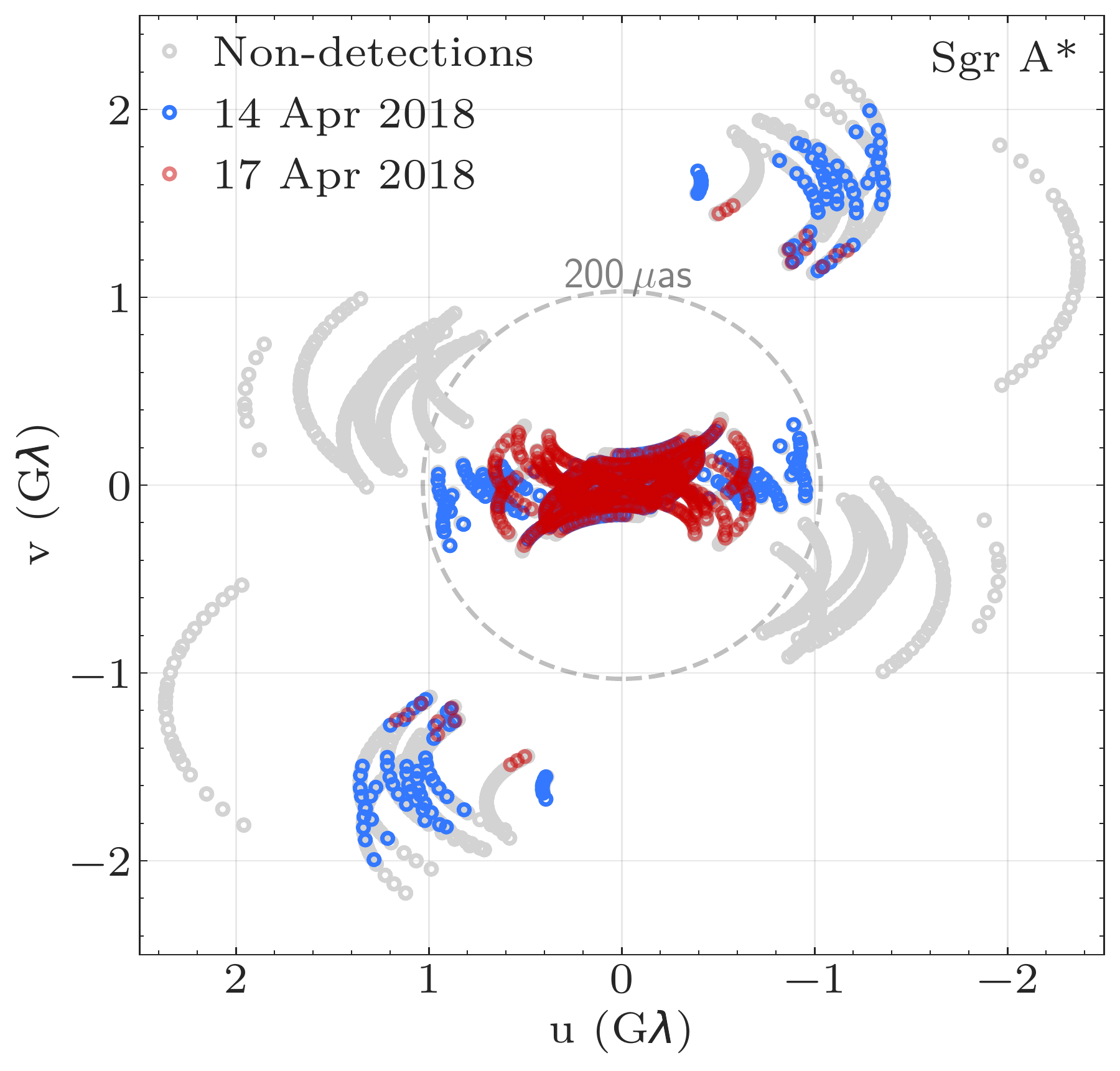}
\includegraphics[width=1.01\linewidth]{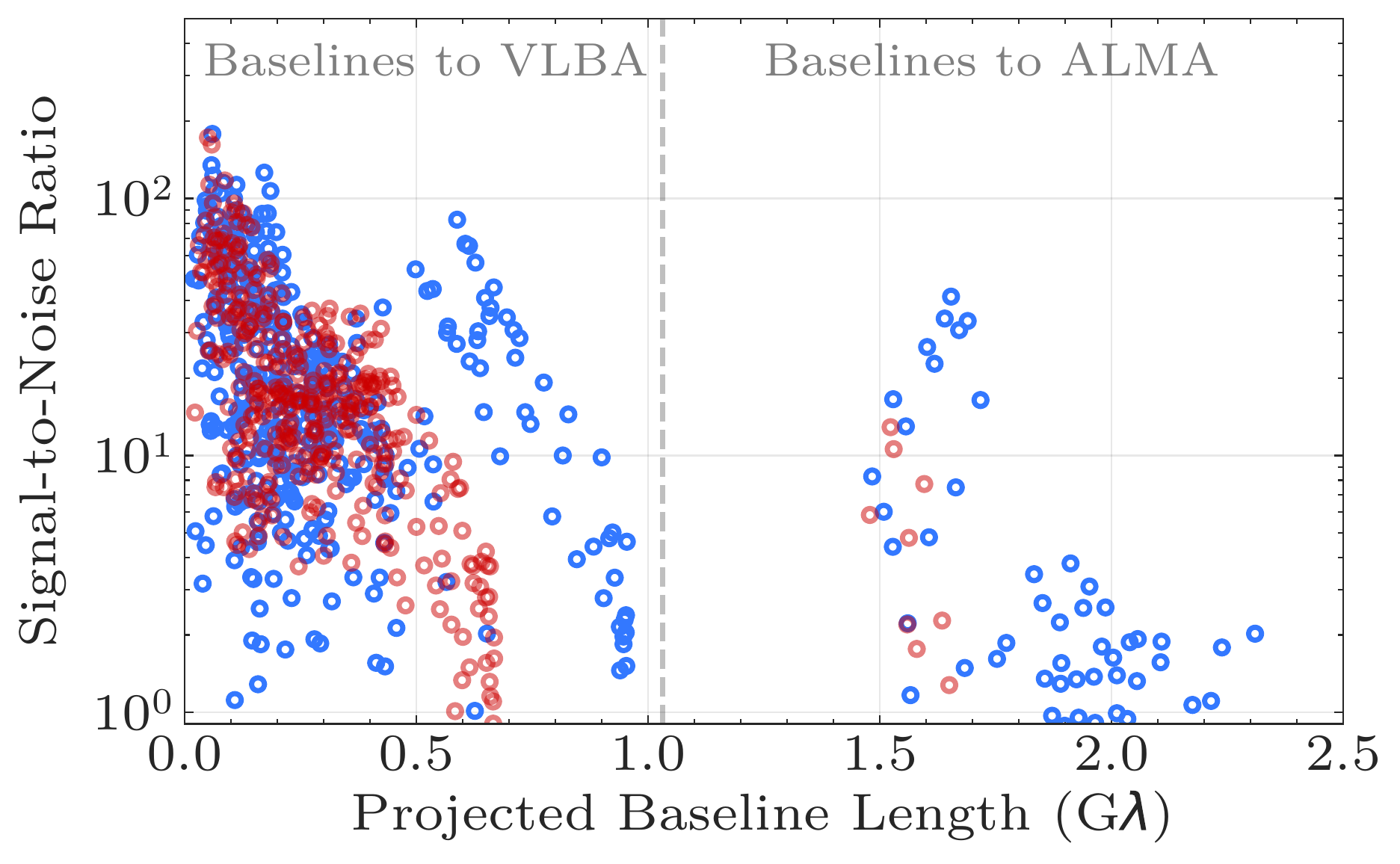}
\caption{{\it Top:} The \sgra\ ($u,v$) coverage, showing non-detections in gray, and detections for 14 April (blue) and 17 April 2018 (red). Each symbol denotes a scan-averaged (over $\sim 9$ minutes) measurement. {\it Bottom:} The signal-to-noise ratio (S/N) for scan-averaged  visibilities on \sgra\ as a function of projected baseline length, showing detections for 14 April 2018 (blue) and 17 April 2018 (red). The gray dashed line in both panels delimits baseline lengths equivalent to a resolution of 200\,$\mu$as. All detections beyond ${\sim}1$\,G$\lambda$ are on baselines to ALMA. 17~April has fewer long-baseline detections due to the absence of the GBT to anchor the fringe calibration of the array, caused by a recording disk failure at the correlation stage. 
}
\label{fig:sgra_cov}
\end{figure}

\section{Observations and data reduction}\label{sec:data}

We observed \sgra\ ($\alpha_\mathrm{J2000} = 17^\mathrm{h}45^\mathrm{m}40^\mathrm{s}.0361$, $\delta_\mathrm{J2000} = -29^\mathrm{\circ}00\mathrm{'}28\mathrm{''}.168$\footnote{Coordinates from the NRAO VLBI observing schedules C181B and C181F are provided by the NRAO SCHED program: \url{http://www.aoc.nrao.edu/~cwalker/sched/Source_Catalog.html}}) with the GMVA (project code MJ001), composed of the eight Very Long Baseline Array (VLBA) antennas equipped with 86\,GHz receivers, the  GBT, and phasing 35 single ALMA antennas \citep{Matthews_2018}. The observations were conducted on 14 and 17~April~2018 as part of the Cycle~5 VLBI session with ALMA (project code 2017.1.00795.V). GMVA stations recorded a total bandwidth of 256\,MHz per polarization divided into eight 32\,MHz-wide intermediate frequencies (IFs), while ALMA recorded overlapping 62.6 MHz IFs separated by 58.59375 MHz that fully covered the GMVA band. The two 6~hr tracks included three calibrator sources: 3C\,279, NRAO\,530, and J1924$-$2914. The GBT participated for 3~hours in each track, but the station recording was lost on 17 April due to a recording disk failure.

The data were correlated with the VLBI correlator at the Max Planck Institute for Radio Astronomy in Bonn using the DiFX software \citep{Deller_2011}. To accommodate the noncongruent IF configuration between ALMA and the other stations, data were correlated over distinct sub-IFs and synthesized back into contiguous GMVA IFs using DiFX tool \texttt{difx2difx}. Mixed linear-circular polarization correlation products between ALMA and the GMVA were transformed to pure circular polarization via \texttt{PolConvert} \citep{Marti_2016}, utilizing a full calibration of the ALMA interferometric products performed by the ALMA quality assurance (QA2) team \citep{Goddi_2018}. Data were then fringe fitted and reduced using the enhanced Haystack Observatory Postprocessing System\footnote{\url{https://www.haystack.mit.edu/tech/vlbi/hops.html}} pipeline ({\tt EHT-HOPS}) presented in \citet{lindyhops}, with additional validation and cross-checks from the NRAO Astronomical Image Processing System \citep[{\tt AIPS};][]{Greisen_2003}. The {\tt EHT-HOPS} pipeline introduces a number of key improvements over the original {\tt HOPS} software, including global fringe fitting and improved phase calibration. Our 86\,GHz implementation of the {\tt EHT-HOPS} reduction follows the same procedure as in \citet{Issaoun_2019}. We performed a priori amplitude calibration and opacity correction with the task {\tt APCAL} within {\tt AIPS}, using observatory-provided telescope gain information and system temperatures measured during the observations. To form Stokes $I$, corrections for field angle rotation and polarimetric gain ratios between left and right polarizations were derived and applied using the EHT Analysis Toolkit\footnote{\url{https://github.com/sao-eht/eat}} (\texttt{eat}
library) polarimetric calibration framework for the {\tt EHT-HOPS} pipeline \citep{lindyhops,Steel2019}.  

Figure~\ref{fig:sgra_cov} shows the detections and non-detections per scan of $\sim 9$ minutes for \sgra\ (top panel)
and corresponding signal-to-noise ratio (S/N) of scan-averaged visibilities for \sgra\ detections for both observed epochs. All detections beyond ${\sim}1$\,G$\lambda$ are on baselines to ALMA. Fringe solutions (delays and delay-rates) are determined from detections with S/N $>$ 7 over the scan, and visibilities on weaker baselines can be measured once station delays and delay-rates are known. After a priori calibration, we proceed to further calibrate antenna gains from calibrator imaging and recover improved measurements of the visibility amplitudes for \sgra.

\section{Results}\label{sec:results}

Sensitivity estimates from a priori amplitude calibration of individual telescopes commonly overestimate station performance, not taking into account effects such as pointing and focus errors, receiver misalignments, operational difficulties, or unstable weather conditions. The visibility amplitudes obtained on \sgra\ from a priori calibration alone do not fully capture true source behavior, and further steps must be taken to disentangle station-based residual amplitude errors from source signal before any more detailed analysis can take place. The amplitude calibration for \sgra\ is done in three stages:
\begin{enumerate}
    \item a priori amplitude calibration using site metadata (see Section~\ref{sec:data});
    \item inner 1\,G$\lambda$ self-calibration to a Gaussian source (see Section~\ref{sec:sgra_cal});
    \item residual self-calibration in \themis modeling of the intrinsic and scattering parameters (see Section~\ref{sec:modeling}).
\end{enumerate}

We present our calibrator imaging in Section~\ref{sec:cal_im}, our calibration methods for \sgra\ data and final gain constraints in Section~\ref{sec:sgra_cal}, and final visibility amplitudes in Section~\ref{sec:sgra_final}.

\subsection{Imaging the Calibrator J1924$-$2914}\label{sec:cal_im}

\begin{figure}[t]
\centering
\includegraphics[width=0.97\linewidth]{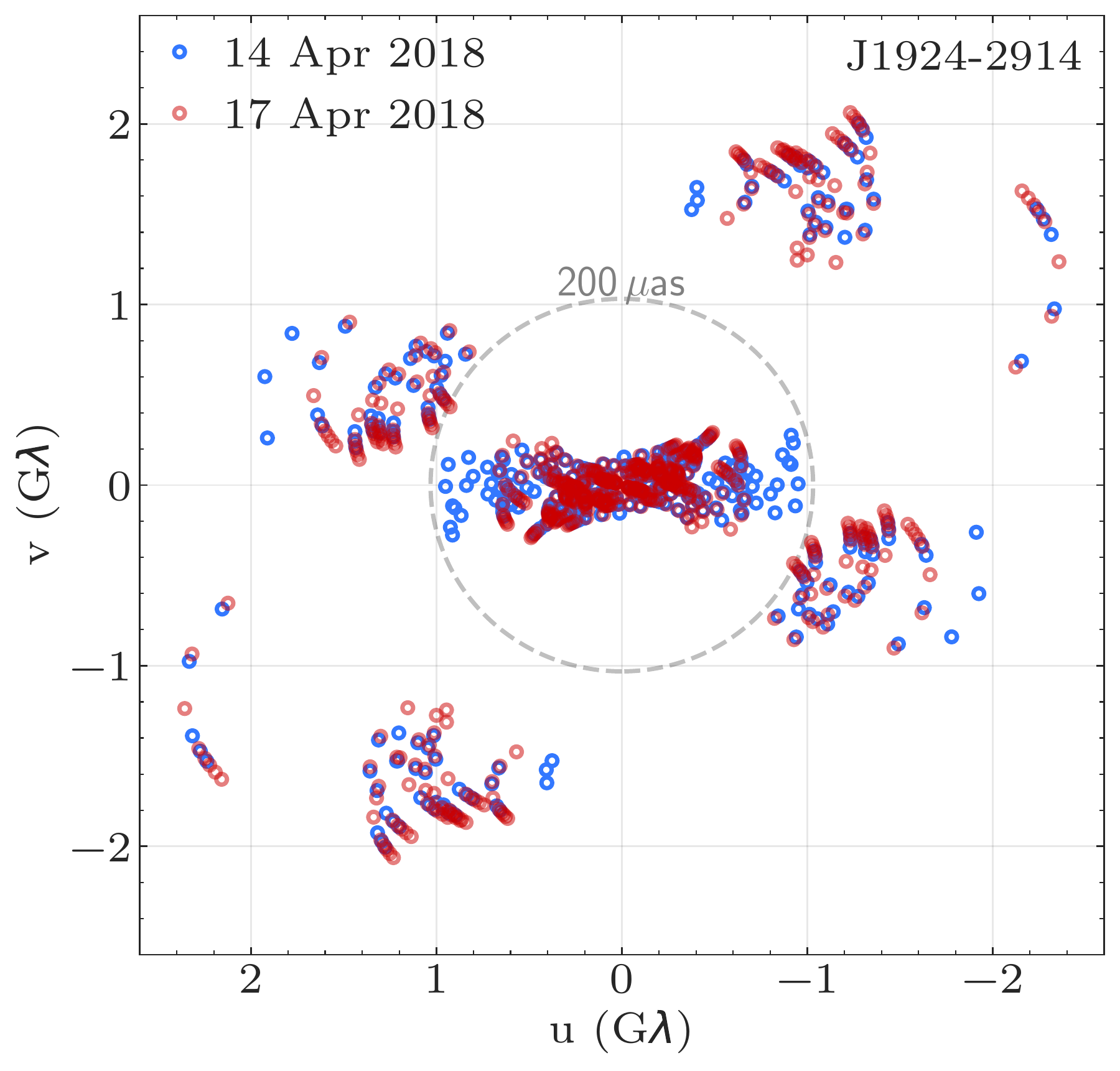}
\caption{The ({\it u,v}) coverage of J1924$-$2914 for both epochs.  14 April scan-averaged detections are shown as blue points, 17 April as red points. The coverage is similar to that of \sgra\  (Figure~\ref{fig:sgra_cov}), but with additional east-west baselines to VLBA Mauna Kea (MK).}
\label{fig:calcov}
\end{figure}

\begin{table}[t]
\caption{Source-integrated flux density of observed sources from interferometric-ALMA.}
\label{tab:fluxes}
\begin{center}
\begin{tabular}{ccc}
\hline 
\hline 
{\bf Source} & {\bf 14 April \textit{\textbf{S$_\nu$}} (Jy)} &  {\bf 17 April \textit{\textbf{S$_\nu$}} (Jy)}  \\
\hline
{\bf \sgra} & $2.2 \pm 0.2$  & $2.3 \pm 0.2$  \\
{\bf NRAO\,530} & $3.2 \pm 0.3$  & $3.2 \pm 0.3$   \\
 {\bf J1924$-$2914} & $4.6 \pm 0.5$  & $4.5 \pm 0.5$   \\
 {\bf 3C\,279} & $14 \pm 1$  & $15 \pm 1$ \\
\hline
\hline
\end{tabular}
\end{center}
NOTE--- These values are provided with interferometric-ALMA data as part of the ALMA Quality Assurance 2 (QA2) process, and assume a $10\%$ uncertainty in the ALMA amplitude calibration \citep{Goddi_2018}. 
\end{table}

\begin{figure*}[t]
\centering
\includegraphics[width=0.274\textwidth]{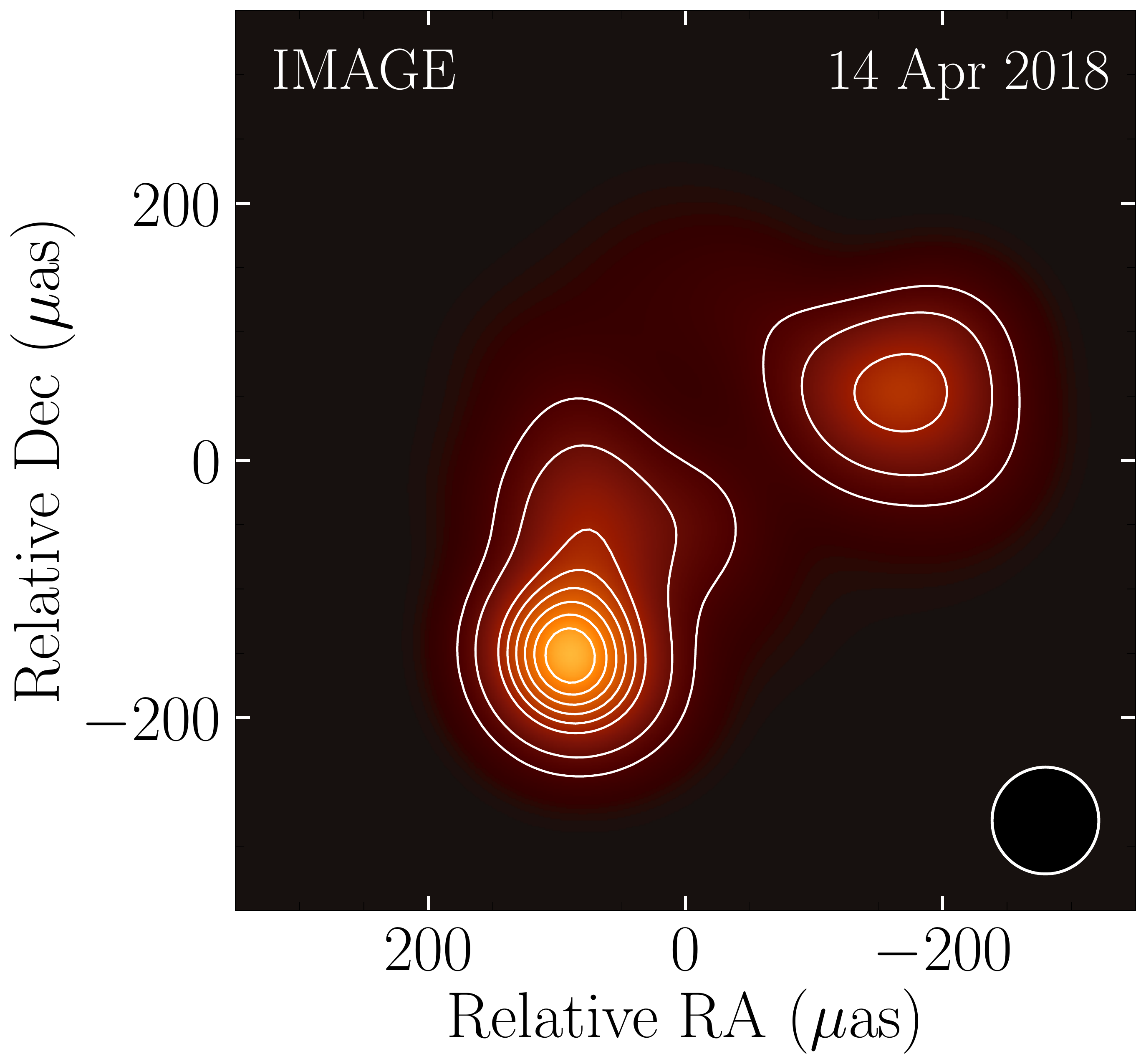}\hspace{-0.05cm}
\includegraphics[width=0.22\textwidth]{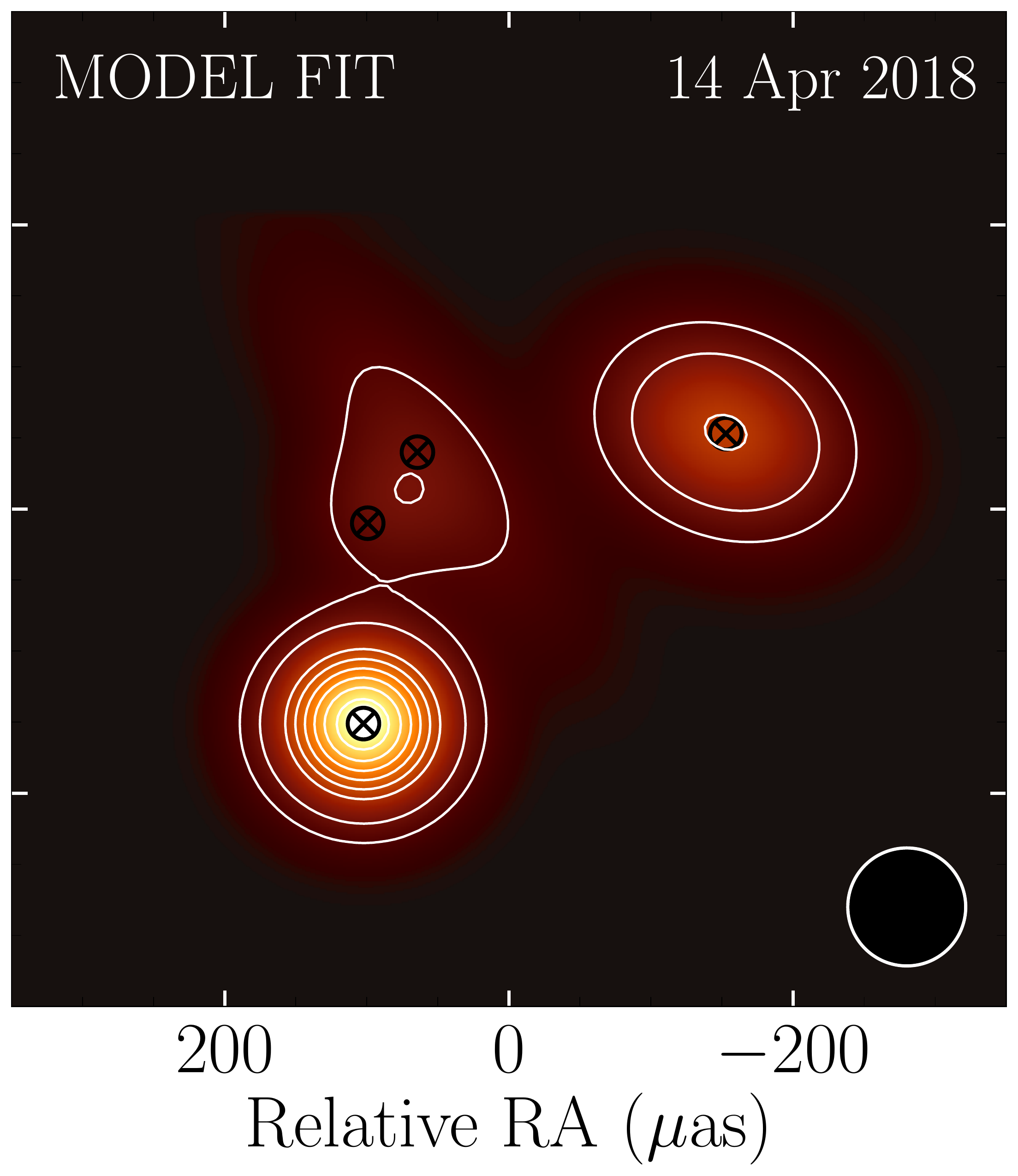}\hspace{-0.05cm}
\includegraphics[width=0.22\textwidth]{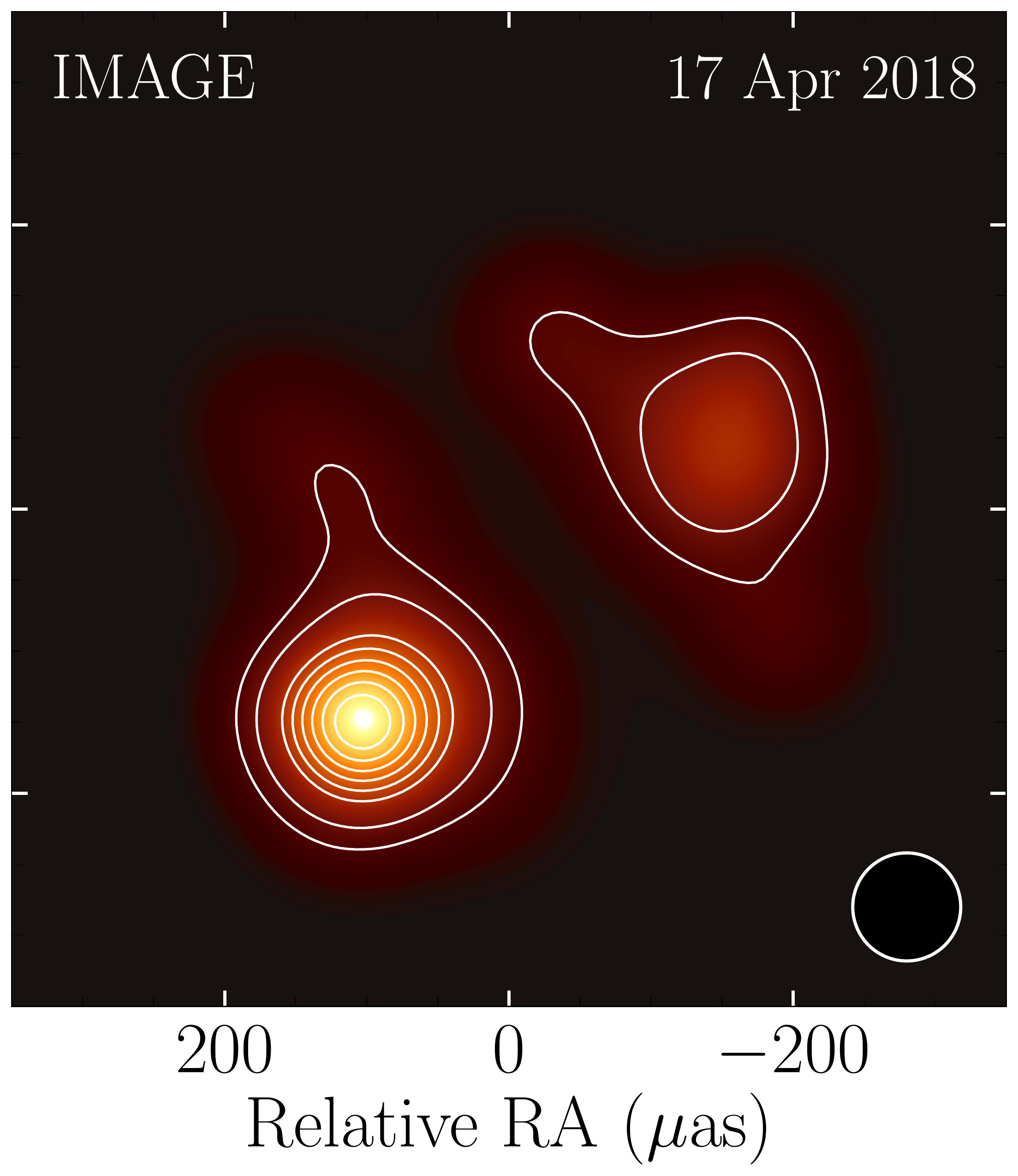}\hspace{-0.05cm}
\includegraphics[width=0.27\textwidth]{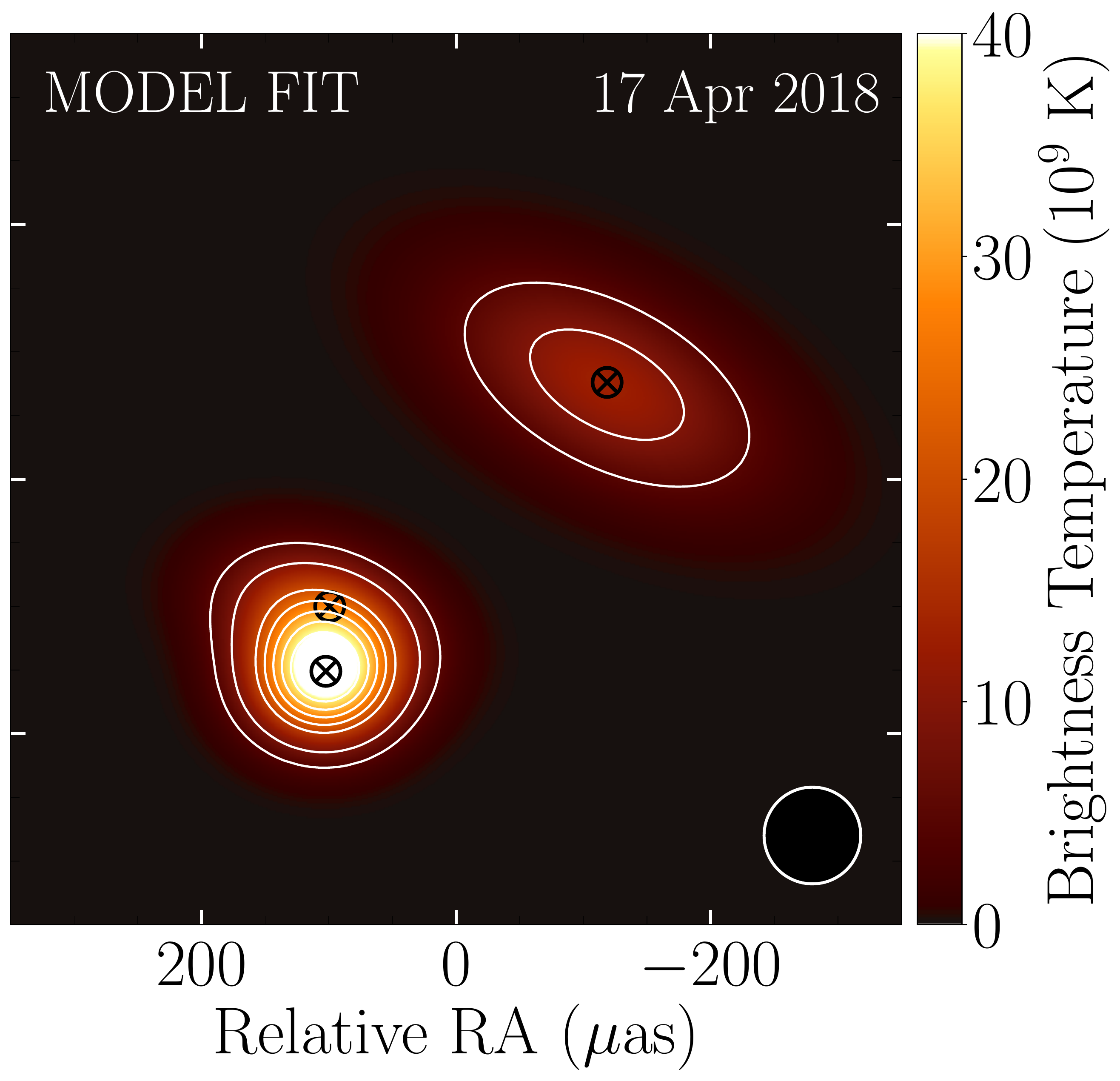}
\caption{ Closure-only images and elliptical Gaussian component model fits of J1924$-$2914 using the {\tt eht-imaging} library. The contour levels are 10--90\% of the peak in steps of 10\%. Black markers denote the central location of each model component in the model fits, produced using the {\tt eht-modeling} module of the library. The images and model fits are restored with a circular Gaussian beam with a FWHM $=1/\mathbf{u}_\mathrm{max} =83\,\mu$as for 14 April and $1/\mathbf{u}_\mathrm{max} =76\,\mu$as for 17 April.
The restoring beam is shown in the lower right corner of each plot. All images and model-fits have reduced $\chi^2<1.7$ on closure products with no systematic noise budget added.}
\label{fig:J1924}
\end{figure*}

To improve the amplitude calibration of the array, we utilize J1924$-$2914, one of the two calibrators observed alternating with \sgra, and with sufficient coverage for imaging (see Figure~\ref{fig:calcov}). J1924$-$2914 appears point-like to ALMA as a connected-element interferometer, and its source-integrated flux density is measured by interferometric-ALMA observations simultaneous to our VLBI tracks (see Table~\ref{tab:fluxes}). Its compactness, stable structure even on VLBI scales within a single epoch, and known flux density make it an ideal imaging target to obtain an estimate of station-based residual gain corrections during our observations via self-calibration. 

While having ($u,v$) coverage comparable to J1924$-$2914, the other imageable calibrator, NRAO\,530, is a north-south extended source \citep{Bower_1997,Bower_1998,Feng_2006,Chen_2010,Lu_2011b,Brinkerink_2019,Issaoun_2019}. In an array configuration where the only north-south baselines resolving the jet are long baselines to ALMA, imaging of the source with the full array proved very difficult, resulting in unphysical images severely impacted by the lack of intermediate baselines between inter-VLBA and ALMA baselines. Imaging without ALMA resulted in various source structures (and varied gain corrections) that provided
statistically acceptable fits to closure quantities, leading to low confidence in any single image structure. We thus omit NRAO\,530 from our gain analysis. We also observed 3C\,279 as a fringe-finder source with the full array for a few minutes but omit it from further analysis because its ($u,v$) coverage is insufficient for imaging.

Following the same method as \citet{Issaoun_2019}, we made use of the large number of closure phases and log closure amplitudes --- constructed via the quotient of two visibility amplitude products involving four stations and robust to station-based errors \citep{Twiss_1960,Readhead1980} --- and the total flux density constrained by the interferometric ALMA measurement in each observation to image J1924--2914 with the {\tt eht-imaging} library \citep{Chael_2016,Chael_2018}.
Closure-only imaging is robust to station-based instrumental errors, allowing us to reconstruct source morphology and derive residual telescope gain corrections directly from self-calibration to the obtained brightness distribution. We confirm general image morphology and gain trends via model fitting to the observed closure products with simple elliptical Gaussian components, and we test goodness-of-fit by calculating a reduced $\chi^2$ of the model prediction against measurements of closure phases and log closure amplitudes.

We present J1924$-$2914 images and model fits\footnote{Throughout this paper we use perceptually uniform colormaps from the \texttt{ehtplot} library, \url{ https://github.com/rndsrc/ehtplot}.} for both epochs in Figure~\ref{fig:J1924}. The observations have a uniform-weighted beam $= (128 \times 83)$ $\mu$as and PA = 38$^\circ$ for 14 April, and a uniform-weighted beam $= (129 \times 76)$ $\mu$as and PA = 45$^\circ$ for 17 April. The resulting images and model fits have reduced $\chi^2 < 1.7$ on closure products with no error budget inflation added to account for possible systematics. A systematic error budget of 1\% (1\% of amplitudes added in quadrature to the thermal noise on complex visibilities) is required to drive reduced $\chi^2$ to unity. The location of the components agree well between the images and model fits of each respective observation. The 14 April morphology is best described with four elliptical Gaussian components based on the lowest reduced $\chi^2$ on the closure data products and realistic station gain reconstructions, while the 17 April morphology, missing GBT baselines, is best constrained with three elliptical Gaussian components. 
On both days, observations indicate a consistent north-west source elongation. The north-west elongation of the J1924$-$2914 jet is consistent with our first image of the source at 3.5\,mm \citep{Issaoun_2019} and follows the mm-jet morphology from 7\,mm \citep{Shen_2002} and 1.3\,mm observations \citep{Lu_2012}.

\begin{table*}[h!t]
\caption{Station median multiplicative gain corrections to the visibility amplitudes for J1924--2914 compared to \sgra.}
\label{tab:gains}
\begin{center}
\begin{tabular}{ccccccc}
\hline 
\hline 
 &  & {\bf 14 April 2018}  & & & {\bf 17 April 2018} & \\
\hline 
{\bf Station} & {\bf \sgra} & {\bf J1924$-$2914} & {\bf J1924$-$2914}  & {\bf \sgra} & {\bf J1924$-$2914} & {\bf J1924$-$2914} \\
 & {\bf Self-Cal} & {\bf Model Fit} & {\bf Image}  & {\bf Self-Cal} & {\bf Model Fit} & {\bf Image} \\
\hline
{\bf ALMA} & --- & 1.8$^{+0.4}_{-0.2}$   & 1.3$^{+0.3}_{-0.2}$ & ---  &  1.1$^{+1.2}_{-0.2}$  &  1.0$^{+1.1}_{-0.2}$   \\
{\bf BR} & 2.4$^{+1.4}_{-1.1}$  & 1.9$^{+2.7}_{-0.9}$   & 2.0$^{+2.7}_{-0.9}$ & 1.6$^{+2.5}_{-0.5}$  &  1.6$^{+2.8}_{-0.5}$  & 1.6$^{+2.7}_{-0.5}$\\
{\bf FD} & 2.2$^{+1.0}_{-0.5}$  & 2.8$^{+2.3}_{-1.1}$   & 2.8$^{+2.3}_{-1.1}$  & 1.9$^{+0.4}_{-0.2}$  &  3.4$^{+3.4}_{-1.9}$  & 3.3$^{+3.1}_{-1.7}$ \\
{\bf GBT} & 1.4$^{+1.6}_{-0.5}$  & 3.1$^{+7.6}_{-1.8}$   &  3.5$^{+5.8}_{-2.4}$ & ---  &  ---  & ---  \\
{\bf KP} & 2.0$^{+1.5}_{-0.7}$  & 2.0$^{+0.7}_{-0.6}$   & 2.0$^{+0.7}_{-0.6}$ & 2.3$^{+1.3}_{-0.7}$  &  2.9$^{+0.6}_{-0.9}$  & 2.9$^{+0.6}_{-1.0}$ \\
{\bf LA} & 1.3$^{+2.1}_{-0.5}$  &  1.2$^{+1.5}_{-0.3}$  & 1.2$^{+1.5}_{-0.2}$ & 1.5$^{+1.3}_{-0.5}$ &  1.3$^{+3.2}_{-0.4}$  & 1.3$^{+3.2}_{-0.4}$ \\
{\bf MK} & ---  & 7.7$^{+2.0}_{-3.7}$   & 5.7$^{+1.0}_{-2.7}$ & --- &  5.3$^{+2.5}_{-1.9}$  & 3.7$^{+1.9}_{-1.1}$ \\
{\bf NL} & 4.7$^{+0.0}_{-0.0}$ & 4.2$^{+1.3}_{-1.7}$   & 4.3$^{+1.3}_{-1.6}$ & 2.3$^{+5.5}_{-0.9}$  &  2.6$^{+2.3}_{-1.2}$  & 2.6$^{+2.3}_{-1.4}$  \\
{\bf OV} & 1.8$^{+2.0}_{-0.3}$  & 1.4$^{+0.4}_{-0.4}$   & 1.4$^{+0.4}_{-0.4}$ & 1.9$^{+2.0}_{-0.3}$  &  1.7$^{+0.4}_{-0.3}$  & 1.8$^{+0.4}_{-0.2}$  \\
{\bf PT} & 2.0$^{+1.4}_{-0.7}$  & 3.6$^{+1.0}_{-1.8}$   & 3.5$^{+1.1}_{-1.8}$ & 2.0$^{+1.2}_{-0.5}$  &  4.0$^{+1.4}_{-1.5}$  &  3.9$^{+1.4}_{-1.5}$ \\
\hline
\end{tabular}
\end{center}
NOTE -- Median multiplicative gain corrections (and 95\% interval of variation over time) to the visibility amplitudes for common stations from the two calibration methods: 1) self-calibration of \sgra\ amplitudes below {1\,}G$\lambda$ to the Gaussian source estimated from \citetalias{Ortiz_2016,Brinkerink_2019}, 2) self-calibration of J1924$-$2914 observations to the closure-only images, and 3) self-calibration of J1924$-$2914 observations to the closure-only model fits.
\end{table*}

\subsection{Calibrating Sagittarius A* Visibility Amplitudes}\label{sec:sgra_cal}

Because of its scatter-broadening, we have fewer detections and lower S/N for \sgra\ than for the calibrators. Our data sets thus do not have sufficiently robust closure quantities to drive the recovery of non-trivial structure in closure-only imaging.
Following the methodology employed in \citet{Issaoun_2019}, we use two methods for amplitude calibration:
\begin{enumerate}
    \item we obtain station gain trends from self-calibration to closure-only model fits and images of a calibrator;
    \item we obtain station gain trends directly from \sgra\ by self-calibrating all visibility amplitudes within 1\,G$\lambda$ using an elliptical Gaussian visibility function obtained from previous 3.5\,mm experiments \citep[hereafter O16, B19 respectively]{Ortiz_2016,Brinkerink_2019}. 
\end{enumerate}

For the second method, we assume that the behavior of the visibility amplitude function for \sgra\ on baselines within 1\,G$\lambda$ is dominated by the image second moment \citep[e.g.,][]{Hu_1962,Issaoun_2019b}, which follows that of the visibility function of a Gaussian source with a FWHM size of 215 by 140\,$\mu$as and a position angle of 80$^\circ$ (east of north). The choice of Gaussian widths is motivated by measurements from previous 3.5\,mm experiments that included the sensitive LMT improving the recovery of the minor axis size \citepalias{Ortiz_2016,Johnson_2018,Brinkerink_2019}. We therefore self-calibrated our \sgra\ amplitudes within 1\,G$\lambda$ to the expected Gaussian morphology and obtain station-based amplitude gain corrections that are subsequently applied to correct visibility amplitudes on all baselines on a time-varying point-by-point basis. Since ALMA does not have any baselines within the 1\,G$\lambda$ cutoff, it cannot be calibrated via this method. 

In Table $\ref{tab:gains}$, we present the median multiplicative station gain corrections obtained via imaging and model fitting of J1924$-$2914 and short-baseline (within 1\,G$\lambda$) self-calibration of \sgra\ to an expected Gaussian source size. The J1924--2914 imaging/modeling and the \sgra\ self-calibration methods gave comparable gain solutions for most stations, validating the Gaussian source assumption for short-baseline measurements of \sgra. According to the VLBA logs, North Liberty (NL) and Mauna Kea (MK) had poor weather on 14 April, which is consistent with the high gain corrections found for these two stations. For 17 April, NL gain corrections are not well constrained as its shortest baseline, to the GBT, is missing. For subsequent analysis, we flag NL for all data sets. MK is not present in the \sgra\ data set, as the source is too scatter-broadened to be detected on long baselines to MK. 
It is worth noting that we tend to recover higher gain corrections from the J1924--2914 Gaussian component model fitting than from the direct imaging for the stations with only long baselines (ALMA, MK), as the model fits do not capture smaller structural variations.
Given the good consistency between imaging and modeling gain corrections for all other stations, we adopt imaging gain corrections for long-baseline stations. Note that we apply all derived gain corrections as a function of time to the data, not just the median scaling presented in Table~\ref{tab:gains}. Because ALMA's multiplicative gain corrections are near unity for J1924--2914, both for imaging and modeling, applying them would not significantly change the flux density on ALMA baselines. We thus choose not to apply ALMA gain corrections as not to introduce scatter from the gain solutions to the visibility amplitudes. 

\begin{figure}[t]
\centering
\includegraphics[width=\linewidth]{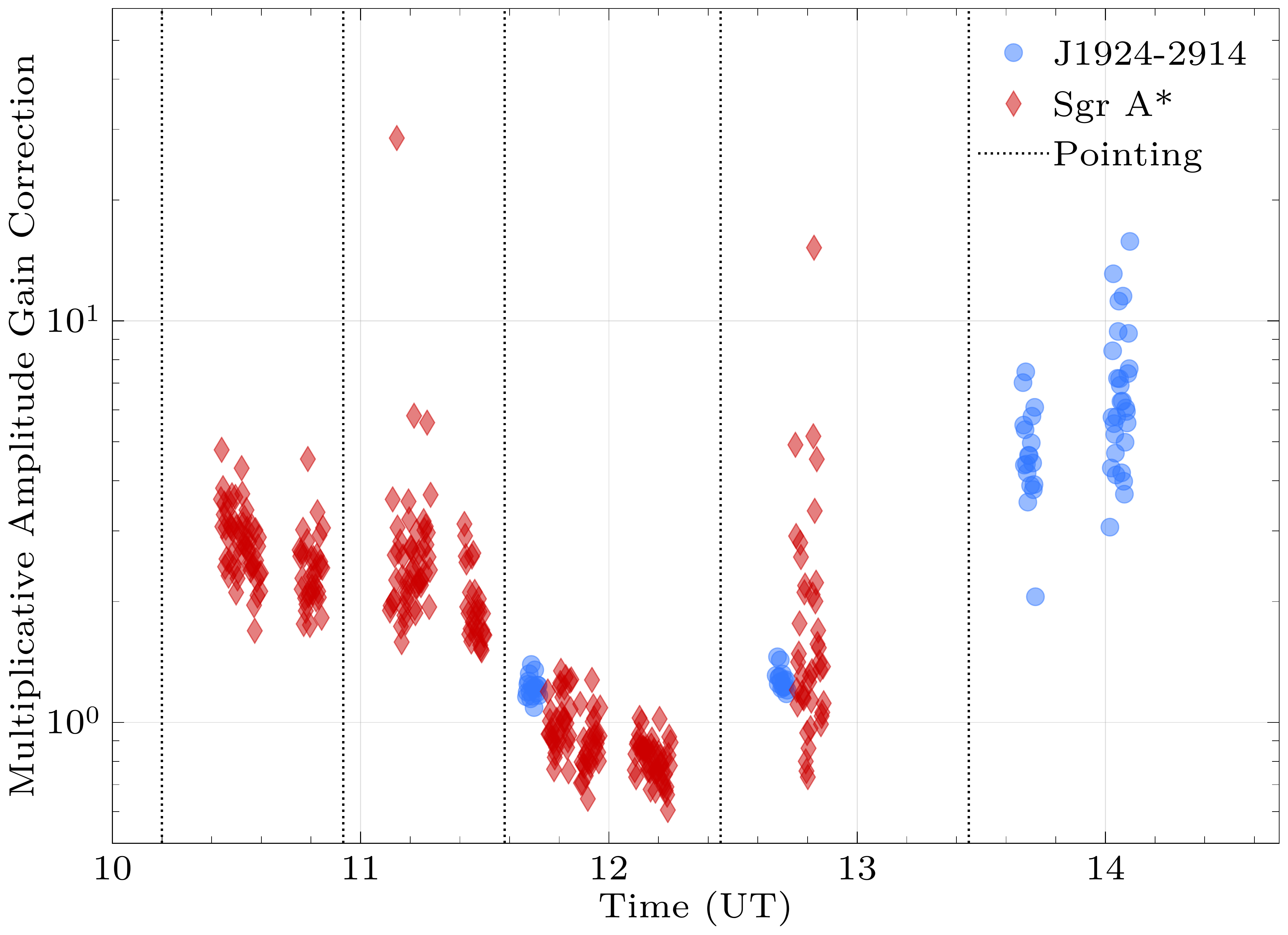}
\caption{GBT multiplicative 10-second interval gain correction trends for J1924--2914 (from the closure-only image) and \sgra\ (from self-calibration to a Gaussian source size) derived from the 14 April observations. The elevations of both sources are descending during the GBT observing track, and remain below $25^\circ$ for all scans. Vertical dotted lines denote the times at which the GBT performed pointing scans on calibrator sources. The intra-scan scatter is likely due to pointing errors. High gain corrections after 13 UT and large scatter are likely due to faulty re-pointing or non-optimal surface adjustment of the telescope, solely affecting half of the J1924--2914 scans.}
\label{fig:gbt_gains}
\end{figure}

\begin{figure*}[h!t]
\centering 
\includegraphics[width=0.9\linewidth]{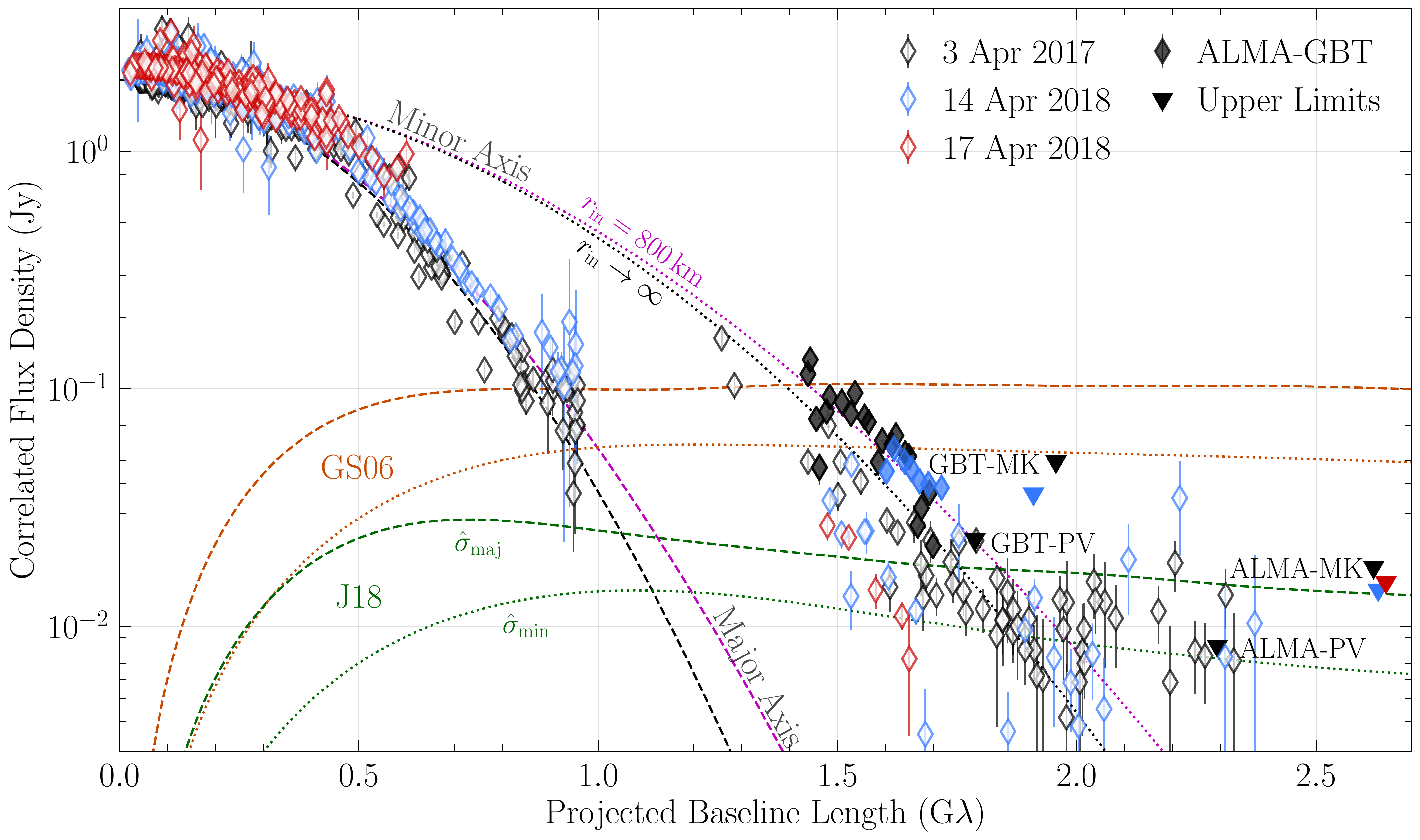}
\caption{Noise-debiased correlated flux density of \sgra\ as a function of projected baseline length for data after self-calibrating to the Gaussian source size from \citetalias{Ortiz_2016} and \citetalias{Brinkerink_2019} using only baselines shorter than $1\,{\rm G}\lambda$. The errorbars indicate the thermal error in individual scans. The amplitude uncertainties from imperfect calibration are not shown but are of the order of $\sim10\%$, corresponding to roughly the size of a symbol. Three epochs are depicted: 3 April 2017 observations presented in \citet{Issaoun_2019} are shown in black; 14 April 2018 observations are shown in blue; and 17 April 2018 observations are shown in red (no GBT). Baselines to NL are flagged for all data sets due to erratic amplitude gain corrections from bad weather. Dashed and dotted curves show the expected visibilities along the major and minor axes, respectively, for an intrinsic elliptical Gaussian source with a FWHM size of $140 \times 105\,\mu$as scattered with two scattering models: in magenta the intrinsic Gaussian source is scattered with the estimated \citetalias{Johnson_2018} scattering parameters $\alpha=1.38$ and $r_\mathrm{in}=800\,$km; while in black the same Gaussian source is scattered with a Gaussian scattering kernel ($169 \times 86\,\mu$as, position angle of $81.9^\circ$ east of north; \citetalias{Johnson_2018}), with $r_\mathrm{in} \rightarrow \infty$. The orange and green curves show the expected RMS renormalized refractive noise along the major and minor axes for the \citetalias{Goldreich_2006} and \citetalias{Johnson_2018} models respectively. All detections beyond ${\sim}1$G$\lambda$ are baselines to ALMA oriented close to the scattering minor axis. Labeled filled triangles, colored by data set, indicate $4\,\sigma$ upper limits on four sensitive baselines at other orientations, where corresponding detections were found for our calibrators.
Detections on the ALMA--GBT baseline (filled diamonds), oriented along the scattering minor axis, sit above the expected flux density for a Gaussian source and thus clearly indicate non-Gaussian source morphology. Detections on baselines beyond 2\,G$\lambda$ exhibit expected properties of refractive noise. For both years, refractive noise detections match the average level predicted by \citetalias{Johnson_2018} and sit below that of \citetalias{Goldreich_2006}.
}
\label{fig:sgra_vis}
\end{figure*}

The stations with significant discrepancies between gain solutions derived from \sgra\ and J1924$-$2914 are GBT and Pie Town (PT) for 14 April, and Fort Davis (FD) and PT for 17 April. Both FD and PT have other VLBA stations very close to them, which allow them to be well-constrained by the Gaussian source assumption for \sgra\, whereas some extended diffuse features may be missing in the imaging/modeling of J1924$-$2914 that lead to this discrepancy. Since the ALMA--GBT baseline is crucial to understanding deviations from the Gaussian source assumption for \sgra, we take a closer look at the derived gain corrections for the GBT using both methods. In Figure~\ref{fig:gbt_gains}, we show the 14 April GBT gain trends for the \sgra\ Gaussian source method and the J1924$-$2914 image as a function of time. The discrepancy between the two sources is due to a systematic offset for half of the GBT scans on J1924$-$2914 at the end of the GBT track. This offset is possibly due to a faulty pointing solution or non-optimal surface adjustment for the telescope after 13 UT affecting half of the J1924$-$2914 scans. For the times where both sources are observed intermittently, there is a good agreement between the derived gain corrections, with mean GBT amplitude gain corrections of $1.25 \pm 0.08$ and $1.2 \pm 0.1$ for J1924--2914 and \sgra\ respectively. We thus choose to proceed with the Gaussian source-derived gain corrections for the calibration of \sgra\ GBT baselines.

\subsection{Final \sgra\ Visibility Amplitudes}\label{sec:sgra_final}

In Figure~\ref{fig:sgra_vis}, we show the scan-averaged noise-debiased visibility amplitudes for \sgra\ after Gaussian source self-calibration of the inner 1\,G$\lambda$ for all GMVA+ALMA observations to date (3 April 2017, 14 April 2018, 17 April 2018). For noise-debiasing, thermal noise contributions to visibility amplitude are removed according to the prescription in \citet{Johnson_2015} \citep[see also][]{TMS}.
The ALMA--GBT baseline, sitting along the scattering minor axis, gives significantly higher flux density than that expected for the minor axis of an intrinsic Gaussian source with a FWHM size of $140 \times 105\,\mu$as scattered by a purely Gaussian scattering screen ($r_\mathrm{in} \rightarrow \infty$; black curves) but matches the expected flux density from an intrinsic Gaussian source of the same angular size scattered with the estimated parameters from the \citetalias{Johnson_2018} scattering model ($r_\mathrm{in} = 800$\,km; magenta curves). In 2018, VLBA detections to ALMA, oriented near the scattering minor axis, show clear deviations from Gaussian behavior for the scattered image of \sgra\ on the sky and exhibit properties expected from refractive noise. We also derive 4$\,\sigma$ upper limits on long baselines with other orientations, based on S/N for detections of the calibrators (filled triangles in Figure~\ref{fig:sgra_vis}). Thus our new observations exhibit similar deviations from Gaussian morphology similar to those of our 2017 observations presented in \citet{Issaoun_2019}. For both 2017 and 2018, the ALMA--GBT baseline exhibits a flux density excess and long-baseline VLBA detections to ALMA are consistent with the average refractive noise predicted by \citetalias{Johnson_2018}, sitting below that of \citetalias{Goldreich_2006}. These two scattering realizations one year apart allow us to confidently exclude \citetalias{Goldreich_2006} as a viable model for the interstellar scattering along our line of sight toward \sgra.

The coverage of our 2018 observations, with only 3 hours including GBT in one epoch, is insufficient to image \sgra\ with the techniques used for the 2017 data set \citep{Issaoun_2019,Issaoun_2019b}. The large amplitude uncertainty on short VLBA baselines and the lack of non-zero closure phases prevent high-fidelity imaging. A wide variety of image morphologies can be obtained with similar goodness-of-fit to the data, therefore any images we obtain would be driven by strong prior assumptions in the imaging process. However, even without imaging, a direct analysis of the visibility amplitudes can provide strong constraints on the intrinsic size of \sgra\ and the inner scale of the interstellar scattering.

\section{Discussion}\label{sec:discussion}

Connecting scattering properties observed at centimeter wavelengths to those at millimeter wavelengths is not trivial, and depends on the relationship between the diffractive scale of the scattering $r_{\rm diff} \sim \lambda/\theta_{\rm scatt}$ and the dissipation scale of turbulence in the scattering material \citep{Psaltis_2018,Johnson_2018}. Despite the precise measurement of the angular broadening size of \sgra\ at centimeter wavelengths, the scattering properties cannot be well constrained without a good understanding of the expected transition to non-$\lambda^2$ and non-Gaussian scattering at millimeter wavelengths, where the dissipation scale is comparable to the diffractive scale. The observations presented in this paper offer a prime opportunity to break the degeneracies between the scattering parameters. 

We assume a single thin scattering screen incorporating observed behavior from long and short wavelength observations. The parameterization of the uncertainties in the scattering properties is motivated with physical models of the ISM material. These models are typically an anisotropic power-law with an index $\alpha$ for the power spectrum governing phase variations, extending between a maximum (or outer) scale $r_{\rm out}$ and a minimum (or inner) scale $r_{\rm in}$. We utilize the scattering model developed by \citet{Psaltis_2018}, assuming the ``dipole" model for the magnetic field wander. The parameters that can be varied are the asymptotic Gaussian source parameters of the scatter-broadening kernel ($\theta_{\rm maj,0}$, $\theta_{\rm min,0}$, and $\phi_{\rm PA}$), the power-law index $\alpha$, and the inner scale of the turbulence $r_{\rm in}$ that cause diffractive kernel deviations from Gaussian and $\lambda^2$ behavior.

In this Section, we fix the asymptotic Gaussian source parameters of the scatter-broadening kernel and the distances between the screen, the observer and \sgra\ to the well-constrained values derived from long-wavelength observations for the \citetalias{Johnson_2018} model. In Section \ref{sec:qualconst}, we provide qualitative constraints on $\alpha$, $r_\mathrm{in}$ and source extent using only long-wavelength constraints and the ALMA--GBT measurements at 3.5\,mm, and in Section~\ref{sec:modeling} we present a full modeling of the scattering and intrinsic source parameters. In Section \ref{sec:powerspec}, we discriminate between scattering models based on the expected and measured power on long baselines in our 3.5\,mm data sets.

\begin{figure}[t]
\centering
\includegraphics[width=\linewidth]{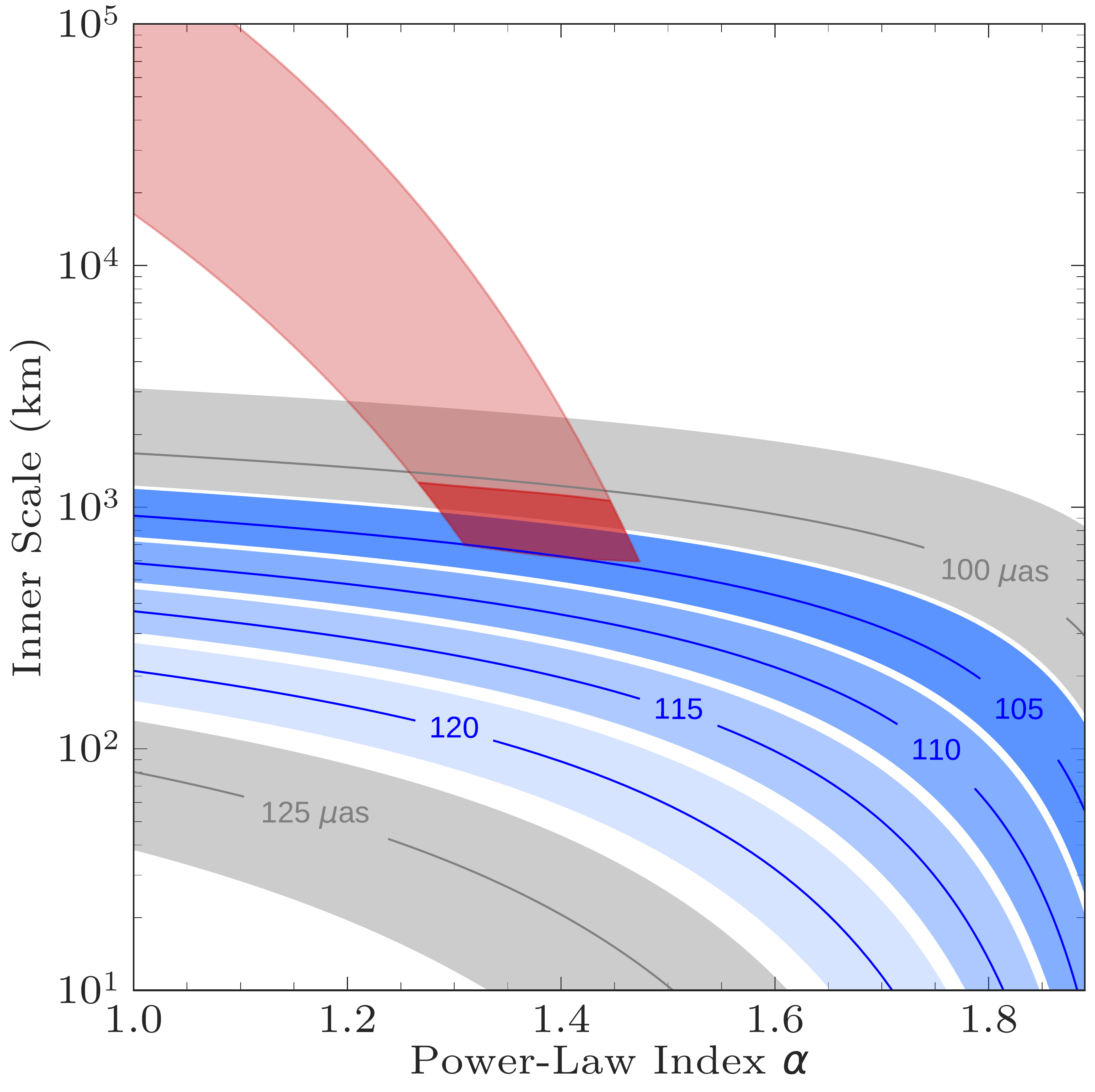}
\caption{Constraints on $\alpha$ and $r_\mathrm{in}$ as a function of intrinsic source extent that would result in the measured mean amplitude ($58 \pm 6$\,mJy) at $\lambda=3.5\,$mm on a projected baseline length of 1.6\,G$\lambda$ along the scattering minor axis. The blue bands show the allowed ranges of $\alpha$ and $r_{\mathrm in}$ for a given intrinsic source size along the scattering minor axis with the measured amplitude at the chosen baseline length. Source sizes with parameter ranges in grayscale lie beyond the allowed intrinsic source FWHM estimates within $1\,\sigma$ (106--121\,$\mu$as) from historical 3.5\,mm measurements \citepalias[\citealt{Issaoun_2019}]{Ortiz_2016,Johnson_2018,Brinkerink_2019}. The red shaded region shows the range of $\alpha$ and $r_\mathrm{in}$ constrained via 1.3\,cm and 7\,mm observations in \citet{Johnson_2018}.}
\label{fig:scatt_sizes}
\end{figure}

\begin{table*}[ht]
\caption{Results from \themis model fitting of the intrinsic source and scattering parameters simultaneously.}
\label{tab:themis}
\begin{center}
\begin{tabular}{ccccc}
\hline 
\hline 
{\bf Variable} & {\bf 3 Apr 2017 } & {\bf 14 Apr 2018} &  {\bf 17 Apr 2018} & {\bf \citetalias{Johnson_2018}}  \\
\hline
$\theta_{\rm min}$ ($\mu$as) & 99$^{+6}_{-7}$ & 96$^{+7}_{-6}$ & 110$^{+40}_{-52}$  & --- \\
$\theta_{\rm maj}$ ($\mu$as) & 146$^{+11}_{-12}$ &  140$^{+14}_{-13}$ &  129$^{+67}_{-20}$ & ---\\
  $\phi$ (deg.) & 70$^{+6}_{-6}$ & 66$^{+16}_{-9}$ &  72$^{+95}_{-63}$ & --- \\
  $\alpha$ & 0.8$^{+0.8}_{-0.7}$ & 1.1$^{+0.8}_{-1.0}$ & 1.0$^{+0.9}_{-0.9}$ & $1.38^{+0.08}_{-0.04}$ \\
 log$_{10}(r_{\rm in}/1\,{\rm km})$ & $3.2^{+0.5}_{-0.6}$ &  $5^{+2}_{-2}$ & $4^{+3}_{-2}$ & $2.9^{+0.1}_{-0.1}$ \\
\hline
\hline
\end{tabular}
\end{center}
NOTE--- Median values and 95\% confidence ranges are shown. The posteriors have been filtered to remove numerical pathologies at the edges of the sampled scattering parameter ranges. 
\end{table*}

\begin{figure*}[ht]
\centering
\includegraphics[width=0.32\linewidth]{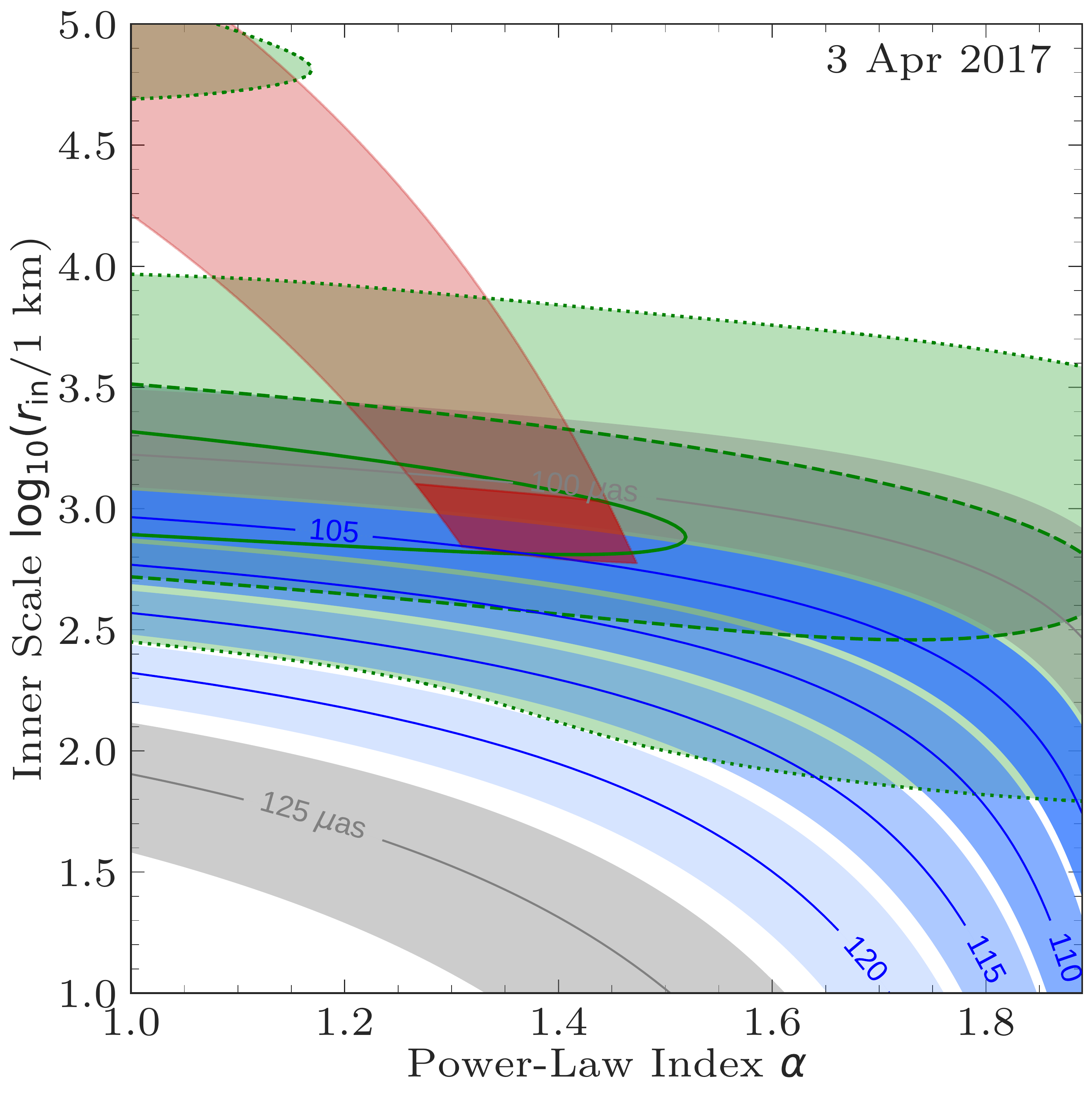}
\includegraphics[width=0.32\linewidth]{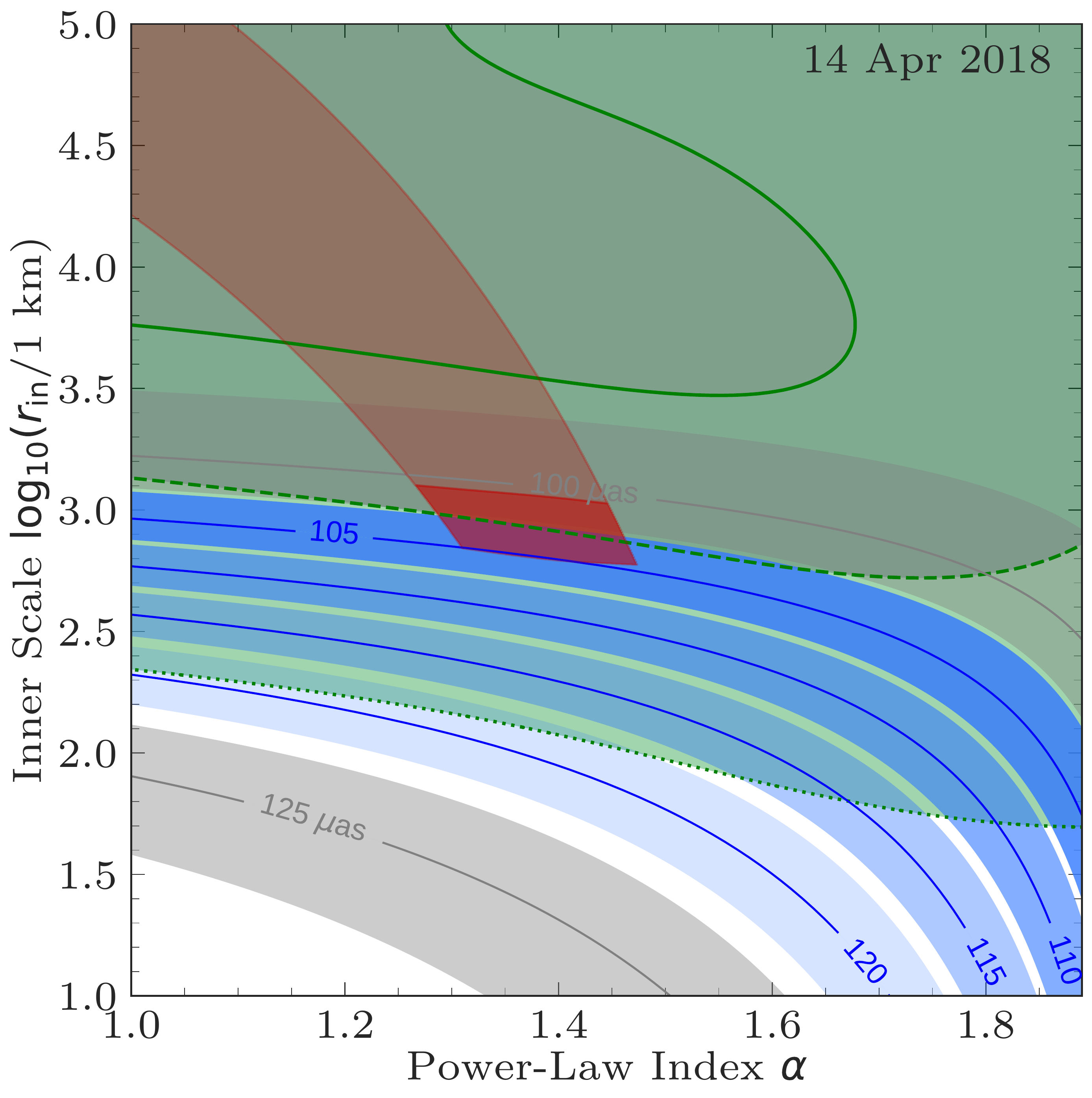}
\includegraphics[width=0.32\linewidth]{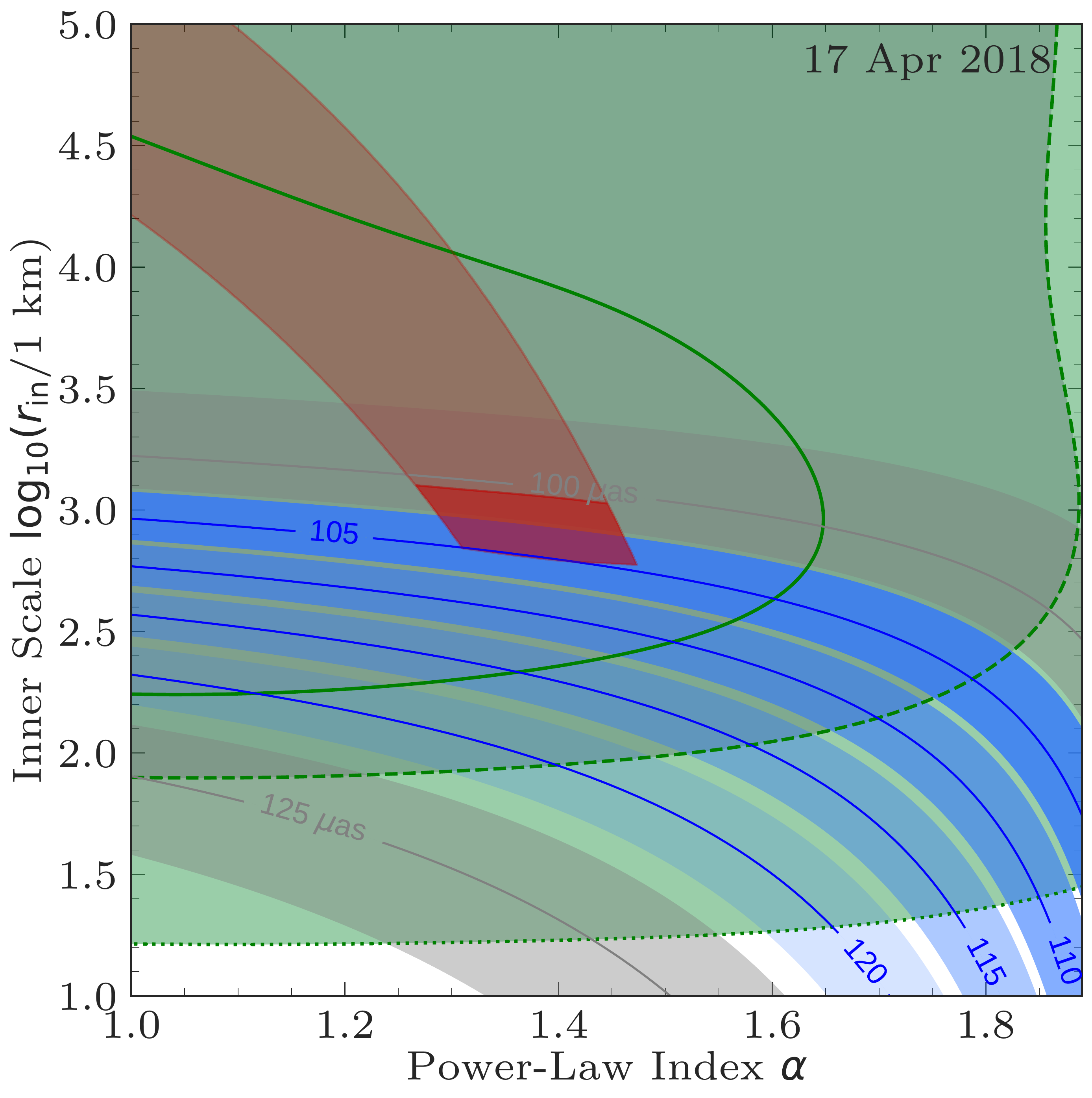}
\caption{Same as Figure~\ref{fig:scatt_sizes}, but with the added \themis posterior distributions from simultaneous intrinsic source and scattering parameter fitting overlaid in green. The 1$\sigma$, 2$\sigma$, and 3$\sigma$ regions are shown from dark green to light green. The solid, dashed, and dotted lines are the  1$\sigma$, 2$\sigma$, and 3$\sigma$ contours respectively.}
\label{fig:scatt_themis}
\end{figure*}

\subsection{Long-Wavelength Constraints on Inner Scale and Intrinsic Size} \label{sec:qualconst}

In Figure~\ref{fig:scatt_sizes}, we present the current constraints on the scattering parameters $\alpha$ and $r_\mathrm{in}$ for various intrinsic source extents derived from 3.5\,mm measurements. 
Our measurements on the ALMA--GBT baseline, oriented along the scattering minor axis and resolving the source, indicate clear non-Gaussian source morphology that is persistent over two years of observations. Our historical size measurements at 3.5\,mm provide constraints for the intrinsic source FWHM along the scattering minor axis to be 106--121\,$\mu$as within $1\,\sigma$ uncertainties assuming the \citetalias{Johnson_2018} scattering model \citep{Johnson_2018,Issaoun_2019}. We measured the combined 2017--2018 mean amplitude in the middle of the ALMA--GBT baseline track to be $58\pm6$\,mJy at a projected baseline length of 1.6\,G$\lambda$, assuming a 10\% uncertainty on the overall amplitude calibration. 

 As shown in Figure~\ref{fig:sgra_vis}, low values for $r_{\rm in}$ lead to a shallower fall-off of the visibility amplitudes as a function of baseline length, while $r_{\rm in}\to \infty$ approaches perfect Gaussian behavior. The same behavior can be achieved with low values of $\alpha$. Therefore, for a measured flux density on a given baseline probing the ensemble-average (purely scatter-broadened) image (with a given intrinsic source size, $\alpha$, and $r_{\rm in}$) the same flux density can be achieved with lower $\alpha$ and $r_{\rm in}$ values paired with a larger intrinsic source. The blue shaded regions in Figure~\ref{fig:scatt_sizes} thus delimit the ranges of parameters giving the measured flux density on ALMA--GBT for a given intrinsic source extent along the scattering minor axis at 3.5\,mm within our $1\,\sigma$ measured range assuming the \citetalias{Johnson_2018} model. The gray shaded regions give examples of source extent that are beyond our $1\,\sigma$ measured range.

The red shaded regions in Figure~\ref{fig:scatt_sizes} delimit composite constraints from longer-wavelength radio observations (see Figure 9 of \citealt{Johnson_2018}). The light red shaded region corresponds to the composite (95\% confidence) ranges of $\alpha$ and $r_\mathrm{in}$ able to reproduce the refractive noise measurements at both $\lambda=3.6$\,cm and $\lambda=1.3$\,cm but unable to reproduce the non-Gaussian source morphology observed at 7\,mm, which requires an inner scale $r_\mathrm{in}<2000$\,km. The dark red shaded region corresponds to the ranges of $\alpha$ and $r_\mathrm{in}$ able to both reproduce the cm-wave refractive noise measurements and the non-Gaussian shape at $\lambda=7$\,mm. The lower limit of the red region corresponds to a lower limit of $r_\mathrm{in}\geq520$\,km constrained by the Gaussian source morphology at 1.3\,cm. Further details on these longer-wavelengths constraints are presented in \citet{Johnson_2018}. From the composite longer-wavelength model constraints and 3.5\,mm size constraints in Figure~\ref{fig:scatt_sizes}, we conclude that the intrinsic source extent (FWHM) of \sgra\ along the scattering minor axis must be $100-105\,\mu$as. Future analysis of more recent longer-wavelength observations of \sgra\ can also further constrain the parameter space of the scattering model (I. Cho et al. \textit{in prep.}). 

\subsection{Joint Modeling of Scattering and Source Parameters} \label{sec:modeling}

Using only the 3.5\,mm observations, we additionally simultaneously model the intrinsic source and interstellar scattering to obtain constraints on source and scattering parameters within the modeling and analysis framework \themis \citep{broderick20}. 
\themis provides a number of methods for handling data, defining models, addressing data systematic uncertainties, and sampling the resulting likelihoods.
Due to the lack of large closure phases, we fit the 3.5\,mm final visibility amplitudes obtained in Section~\ref{sec:sgra_final}, excluding those data points with an S/N less than 2 to avoid non-Gaussian errors \citep{TMS}.

The primary output of \themis-based analyses are posteriors on model parameters implied by the input data.  In this instance, to ensure global convergence, we use the parallelly-tempered Markov Chain Monte Carlo (MCMC) sampler: we employed the deterministic even-odd swap tempering scheme of \citet{DEO:2019} with the automated factor slice sampler of \citet{AFSS:2014}, which is very efficient for small numbers of model parameters.

\sgra\ is modeled as an 
elliptical Gaussian source convolved with a parameterized version of the asymmetric diffractive kernel described in \citet{Psaltis_2018} and \citet{Johnson_2018}.  The 
elliptical Gaussian source is parameterized in terms of a total flux density, averaged size, asymmetry parameter (major-to-minor axis ratio), and position angle ($\phi$) as described in \citet{broderick20}, each with uninformative uniform priors.  From these we construct the intrinsic source Gaussian major/minor axes ($\theta_{\rm maj},\theta_{\rm min}$).  Only the inner scale $r_{\rm in}$ and power-law index $\alpha$ are permitted to vary, with the remaining scattering parameters already well constrained by prior data.  For $r_{\rm in}$ a logarithmic prior is assumed, ranging from 1\,km to $10^7$\,km; for $\alpha$ a uniform prior is assumed on $(0,2)$. 

\begin{figure*}[t]
\centering
\themis gain corrections before inner 1\,G$\lambda$ self-cal
\includegraphics[width=\linewidth]{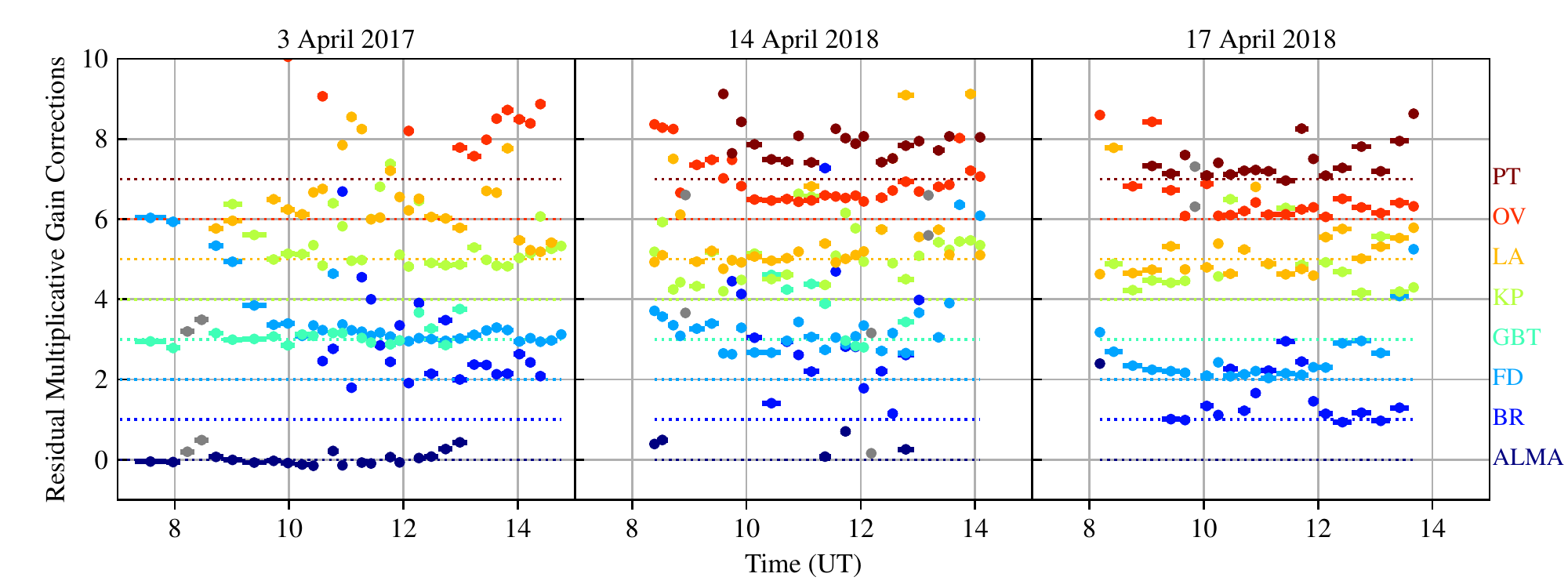}
\themis gain corrections after inner 1\,G$\lambda$ self-cal
\includegraphics[width=\linewidth]{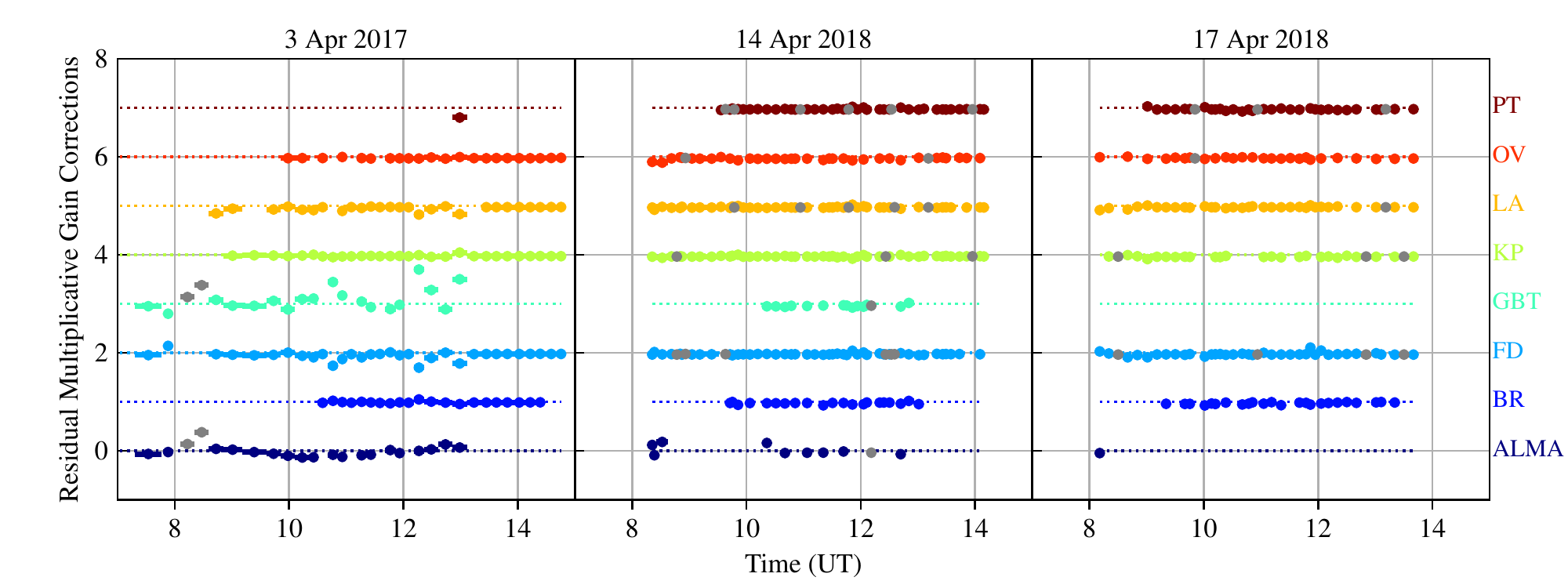}
\caption{Residual multiplicative gain corrections derived by \themis in the \sgra\ model fitting for all three observing days. {\it Top: Residual variations in the amplitudes derived from the data sets where only a priori amplitude corrections are applied. {\it Bottom}:} Residual variations in the amplitudes derived from the data sets already corrected based on self-calibration (see Section~\ref{sec:sgra_cal}). For 3 April 2017, inner 1\,G$\lambda$ self-cal gain corrections were not applied to the GBT due to the scatter introduced by its baseline to NL, whose measurements dominate within that baseline cut \citep{Issaoun_2019}. In all panels, individual station gain corrections are offset vertically by unity for clarity,  with the dashed horizontal lines indicating a unit gain for each station. The gain correction scale is linear. Each point corresponds to an individual scan: colored points are independently reconstructed in the procedure, grey points are not well constrained due to missing information and are thus heavily biased by the prior. }
\label{fig:residual-gains}
\end{figure*}

To accommodate the refractive scattering component a constant 7\,mJy noise floor is added in quadrature to the input data uncertainties; this is both conceptually and computationally much simpler than fully modeling the complex phase screen. Station residual multiplicative gain corrections are reconstructed and marginalized over via the Laplace approximation during the construction of log-likelihoods, after imposing a Gaussian prior centered on unity with a standard deviation of 20\% \citep[see Section 6.8 of][]{broderick20}. 
Fitting gain amplitudes has an added practical benefit: when fitting mock data sets, produced with similar noise properties and baseline coverage to our observations of \sgra, we found that fitting the gain amplitudes effectively mitigated biases from refractive scattering.
The derived residual multiplicative gain corrections are shown in Figure~\ref{fig:residual-gains}.
Each data set was independently analyzed to avoid potential complications associated with source variation. Excellent fits were obtained in all cases.  No qualitative differences were found when analyzing both 2018 data sets together.

In Figure~\ref{fig:scatt_themis}, we present the posteriors on the scattering parameters $r_{\rm in}$ and $\alpha$ added to the multi-wavelength analysis from Figure~\ref{fig:scatt_sizes}. The median values and 95\% confidence ranges for all fitted parameters are presented in Table~\ref{tab:themis} for each data set. The results for 3 April 2017 and 14 April 2018 show great consistency within our confidence ranges, while the results for 17 April 2018 are not well-constrained due to the absence of the GBT. Although $\alpha$ is not well-constrained at 3.5\,mm, the 2017 data set in particular constrains $r_{\rm in}<3300\,$km at the 2$\sigma$ level. The 2017 data set contains the longest observation with the ALMA--GBT baseline, on which the amplitude fall-off indicates deviations from Gaussian morphology and thus results in a finite inner scale, as illustrated in Figure~\ref{fig:sgra_vis}. This result is completely independent from the longer-wavelength constraints (red shaded regions in Figure~\ref{fig:scatt_themis}), yet provides complete overlap in the parameter space for $\alpha$ and $r_{\rm in}$, as well as a consistent intrinsic size estimate to that derived from longer-wavelength constraints in Section~\ref{sec:qualconst}. 
While the 2018 data sets alone do not constrain the $\alpha-r_{\rm in}$ parameter space, the clear ALMA--GBT detections on 14 April 2018 build confidence on the non-Gaussian behavior on that baseline across multiple years. The persistently low-flux-density long-baseline VLBA--ALMA detections allow us to rule out the \citetalias{Goldreich_2006} scattering model as a viable model for the interstellar scattering toward \sgra, see Section~\ref{sec:powerspec}.

We select the 3 April 2017 data set as our best data set, due to its more complete $(u,v)$ coverage and its longer GBT observing track (double that of the 2018 observing days).
We obtain the following source size parameters for our best data set (2017): a major axis FWHM of $146 \pm 12\,\mu$as; a minor axis FWHM of $99 \pm 7\,\mu$as; and a position angle of $70\pm 6^\circ$, almost oriented along the diffractive kernel. 
This alignment may be coincidental, or it may indicate that the assumed diffractive kernel is incorrect, producing a biased intrinsic size that is aligned with the scattering kernel. Because our imaging of the 2017 data set in \citet{Issaoun_2019} yielded a largely unconstrained position angle, this alignment may also indicate that the Gaussian model for intrinsic structure is overly simplistic, with tight posteriors that are spurious.
These parameters give a major-to-minor axis ratio (or ``axial ratio") for the source of $1.5 \pm 0.2$. This value is slightly larger than the ratio ($1.2^{+0.3}_{-0.2}$) measured directly from the intrinsic image reconstructed by \citet{Issaoun_2019}. In addition to different systematics due to varying methods for the intrinsic size measurement, the $\alpha$ and $r_\mathrm{in}$ parameters used to separate the scattering effects from the intrinsic source structure are different, shown in Table~\ref{tab:themis}. In \citet{Issaoun_2019} we assumed the \citetalias{Johnson_2018} parameter values, whereas in this work we fit for $\alpha$ and $r_\mathrm{in}$ within \themis simultaneously with the intrinsic source parameters. Thus, the degeneracy in the scattering and intrinsic source size parameters likely lead to this slight shift in the measurement for the axial ratio. Whereas in \citet{Issaoun_2019} we assumed particular values of $\alpha$ and $r_\mathrm{in}$ to obtain information on the intrinsic source structure, our new result, however, confirms that the source still modestly deviates from circular symmetry even if we allow the scattering parameters to vary. 

In \citet{Issaoun_2019} we explored a set of general-relativistic magnetohydrodynamic (GRMHD) simulations of the \sgra\ accretion flow: disk versus jet driven emission, varying particle acceleration, varying spin, and varying heating prescriptions \citep{Moscibrodzka_2009,Moscibrodzka_2014,Moscibrodzka_2016,Davelaar_2018,Howes_2010,Rowan_2017,Chael_2018b}. We found that high-inclination jet-dominated models ($>20^\circ$ of face-on) produce larger axial ratios than the measurement on \sgra, and are thus ruled out. Our new axial ratio measurement in this work, allowing the freedom to fit the scattering parameters, confirms this result. This range in allowed inclination for jet models is also consistent with the independent near-infrared orbiting flare results from the Gravity experiment \citep{Gravity_2018b}. 
In addition, the measured axial ratio is now slightly too large to also be consistent with low-inclination disk-dominated models ($<20^\circ$), which are highly symmetrical \citep[Figures 9 and 10 of][]{Issaoun_2019}. With this new measurement, the explored emission models best able to replicate the size and axial ratio observed for \sgra\ are thus low-inclination jet or mid/high-inclination disk models. However it is worth noting that the degeneracy in the scattering and intrinsic source parameters can be the cause of this shift in axial ratio measurement, and only a small subset of GRMHD models were studied. Therefore whether low-inclination disks are truly ruled out remains uncertain.

\begin{figure}[t]
\centering
\includegraphics[width=\linewidth]{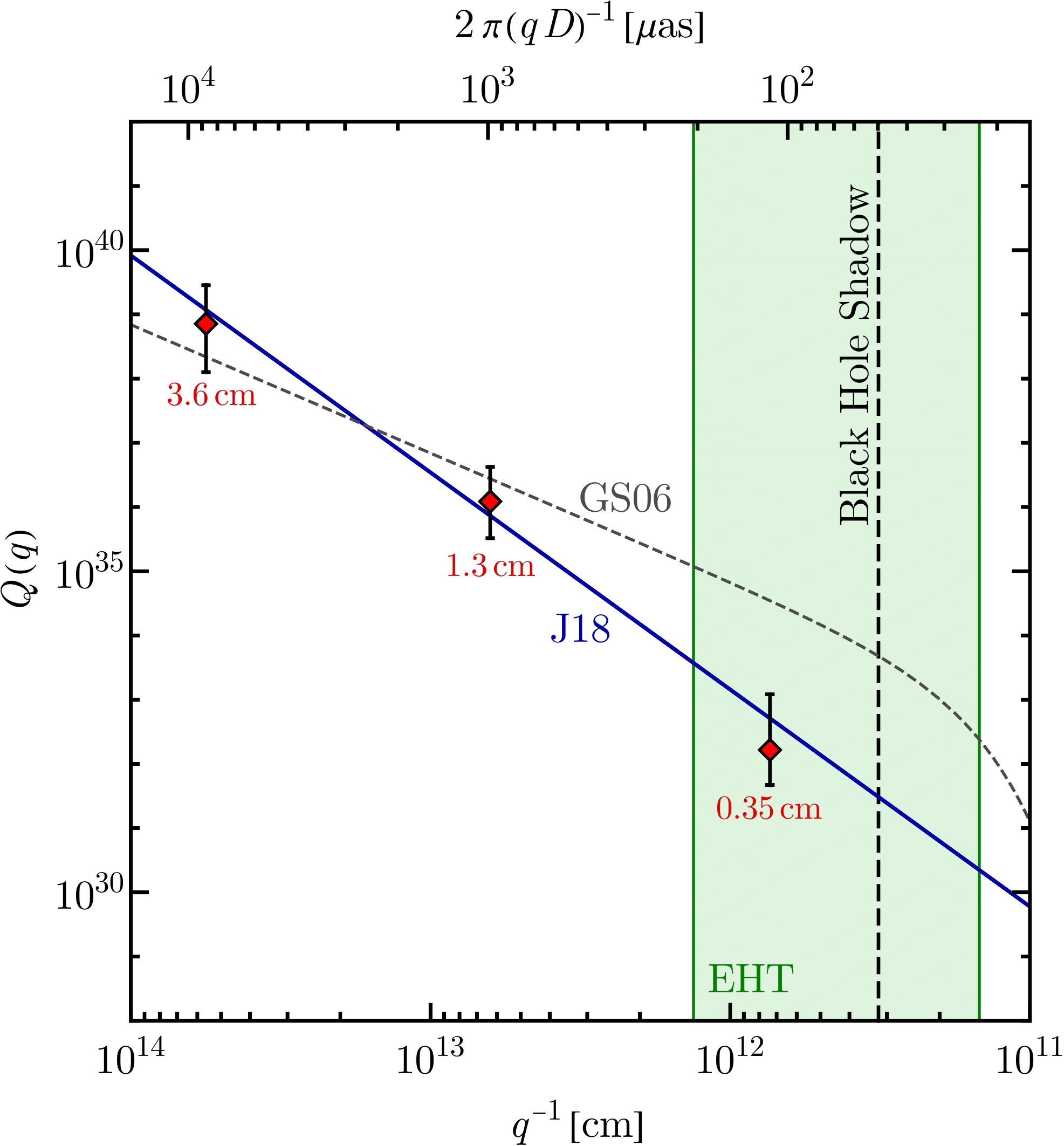}
\caption{Observational constraints, from refractive noise on long interferometric baselines, on the power spectrum $\mathcal{Q}(\mathrm{\bf q})$ of phase fluctuations. Note that $\mathcal{Q}(\mathrm{\bf q})$ is dimensionless and independent of the observing wavelength.
Constraints at 3.6 and 1.3\,cm are from \citetalias{Johnson_2018}; the 3.5\,mm results are from the observations reported here and by \citet{Issaoun_2019}. The red diamonds represent constraints on the power for wavenumbers $q^{-1}\sim 10^{12}-10^{14}$\,cm from refractive noise on long baselines observed at 3.6\,cm, 1.3\,cm and new 3.5\,mm wavelengths along the scattering major axis. The green shaded region delimits the range of modes that are expected to contribute refractive noise to 1.3\,mm EHT images of \sgra. The power spectra of two scattering models are plotted: the \citetalias{Goldreich_2006} model with $\alpha=0$ and $r_\mathrm{in} = 2\times10^6$\,km (dashed gray); and the recommended \citetalias{Johnson_2018} model with $\alpha=1.38$ and $r_\mathrm{in} = 800$\,km (solid blue). All observational constraints are consistent with the recommended \citetalias{Johnson_2018} model. While the previous cm wavelength detections of refractive noise are consistent with both models, the new 3.5\,mm upper limit for the power from refractive noise along the scattering major axis is a factor of 10 below the predicted power by the \citetalias{Goldreich_2006} model.}
\label{fig:Qconstraints}
\end{figure}

\subsection{Constraints on the Power Spectrum} \label{sec:powerspec} 

The refractive noise power on a long baseline $\mathbf{u}$ is chromatic, dependent on the power-law index $\alpha$ of the phase fluctuations, and is dominated by refractive modes with $q \sim 2\pi \mathbf{u}/ D$. Previous constraints on the power spectrum $\mathcal{Q}(\mathbf{q})$ for a wavenumber $q^{-1}$ based on cm-wavelength refractive noise power detections by \citet{Johnson_2018} showed that both the \citetalias{Johnson_2018} and the \citetalias{Goldreich_2006} models fit cm-wave constraints but the models are expected to be most different in the mm-wave regime. In our 2017 and 2018 data sets, we solely detect power on long baselines on ALMA baselines sampling mostly along the scattering minor axis. However, previous cm-wave measurements of noise power were along the scattering major axis \citep[see][Figure 14]{Johnson_2018}. To add a 3.5\,mm constraint to those measurements, we therefore need to make use of estimated $4\,\sigma$ upper limits on baselines along the scattering major axis where no detections were found on \sgra\ but the level of noise for those baselines is known from detections on calibrators.

In Figure~\ref{fig:Qconstraints}, we present a similar plot to Figure 14 in \citet{Johnson_2018}, with our newly added 3.5\,mm constraint. Our mean visibility amplitude (after noise-debiasing) is 6\,mJy along the scattering minor axis on baselines beyond 1.8\,G$\lambda$, measured on ALMA baselines for our 2017--2018 data sets. We determine the central value (red diamond) for the major axis constraint as the equivalent power measured along the scattering major axis at 3.5\,mm for a scattering model giving 6\,mJy of refractive noise along the minor axis. The upper limit of the constraint is given by the $4\,\sigma$ upper limit of 30\,mJy at 1.8\,G$\lambda$ on the sensitive east-west GBT--IRAM 30\,m baseline from our 2017 data set \citep{Issaoun_2019}. The lower limit of the constraint assumes that the 6\,mJy detection for a single scattering realization is purely refractive noise, giving a lower-limit of $\sim$3\,mJy for the ensemble-average RMS \citep[see Section 6.1 of][]{Issaoun_2019}. Because we are combining estimates of refractive noise along the major and minor axes, this estimate of $Q(\mathbf{q})$ is sensitive to the assumed model of magnetic field wander \citep[here, we use the ``dipole'' model of][]{Psaltis_2018} but primarily depends on the rough extent of the ensemble-average image so is insensitive to the assumed scattering parameters, $\alpha$ and $r_{\rm in}$ \citep[see, e.g., Eq.~16 of][]{Johnson_Narayan_2016}. 
 
We also plot in Figure~\ref{fig:Qconstraints} the power spectra for the two scattering models, \citetalias{Goldreich_2006} (dashed gray line) and \citetalias{Johnson_2018} (blue solid line). Refractive modes impacting horizon-scale reconstructions with the EHT are within the green shaded region. While both models fit constraints in the cm-wave regime, the 3.5\,mm constraint discriminates between the two: the power predicted by the \citetalias{Johnson_2018} model is consistent with the constraint, while the power predicted by \citetalias{Goldreich_2006} model lies an order of magnitude above it. The 3.5\,mm measurement directly probes refractive modes within the EHT's field of view, for which scattering substructure can contaminate EHT horizon-scale images. Assuming the \citetalias{Johnson_2018} model, our 3.5\,mm result confirms expectations of low (of the order $\sim1\%$ of total flux density) levels of refractive noise on long VLBI baselines observable with the EHT at 1.3\,mm.

\section{Summary}\label{sec:summary}

We have presented 2018 observations of \sgra\ using ALMA in concert with the GMVA at 86\,GHz. In combination with observations carried out in 2017 presented in \citet{Issaoun_2019}, we show that the ALMA--GBT baseline resolves persistent non-Gaussian morphology along the scattering minor axis. In addition, long-baseline detections to ALMA (1.8--2.4\,G$\lambda$) exhibit characteristics of low-level refractive noise across both years of observations. Using the scattering model developed by \citet{Psaltis_2018}, we show that these long-baseline detections are consistent with the scattering parameters estimated in \citet{Johnson_2018}, while they are at least five times weaker than the expected refractive noise for the parameters suggested by \citet{Goldreich_2006}. While a single realization of the refractive scattering would give values this low for approximately 4\% of observations, the probability of seeing two independent observations at this level is less than 0.15\%, firmly ruling out the \citet{Goldreich_2006} parameters.

We made use of the chromatic nature of the interstellar scattering to put stringent constraints on the intrinsic source extent at 86\,GHz along the scattering minor axis. We combined 8, 22, and 43\,GHz constraints on the scattering parameters presented in \citet{Johnson_2018} with our 86\,GHz persistent flux density excess on the ALMA--GBT baseline. We found that the ALMA--GBT baseline flux density excess predicts an intrinsic source extent along the scattering minor axis of $\sim 100\,\mu$as for \sgra\ for the ranges of power spectrum power-law index $\alpha$ and inner scale of turbulence $r_{\rm in}$ allowed by lower-frequency measurements from \citet{Johnson_2018}. 

Direct modeling of our 86\,GHz data with \themis \citep{broderick20}, fitting simultaneously the scattering and intrinsic source parameters, gave overlapping $\alpha-r_{\rm in}$ parameter ranges with lower-frequency constraints and source size estimates consistent with the lower-frequency predictions. Source size estimates for all independent data sets are consistent within their uncertainties. The fitted source size parameters obtained for our best data set, on 3 April 2017, are: a major axis FWHM of $146 \pm 12\,\mu$as; a minor axis FWHM of $99 \pm 7\,\mu$as; and a position angle of $70\pm 6^\circ$, almost oriented along the diffractive kernel. We obtain a major-to-minor axis ratio for the source of $1.5 \pm 0.2$. In the scope of the set of general-relativistic magnetohydrodynamic simulations explored in \citet{Issaoun_2019}, high-inclination jet-dominated models produce larger axial ratios are ruled out, but the measured axial ratio is now too large to also be consistent with low-inclination disk-dominated models \citep[Figures 9 and 10 of][]{Issaoun_2019}. Assuming the fitted scattering parameters, the explored emission models best able to replicate the size and axial ratio observed for \sgra\ are thus low-inclination jet or mid/high-inclination disk models.

We have shown that 86\,GHz measurements of the interstellar scattering are crucial for the discrimination of scattering models. In particular, our observations favor a dissipation scale of ${\sim}\,10^3\,{\rm km}$ for interstellar turbulence, which is consistent with a characteristic scale determined by the ion Larmor radius \citep{Spangler_Gwinn_1990}. Irrespective of the specific scattering properties, our results probe the scales of phase fluctuations that are comparable to the angular size of the black hole shadow of \sgra, giving model-independent insights into how scattering may affect images of \sgra\ produced with the Event Horizon Telescope.
Future observations at 86\,GHz with the newly upgraded 50-m surface of the LMT will enable probing of the structure along the scattering minor axis on a highly sensitive baseline with ALMA, providing a second anchor for deviations from Gaussian morphology and model constraints. GMVA bandwidth enhancements and station expansion would also enable higher fidelity imaging, enable higher sensitivity in the east--west to possibly detect refractive structure along the scattering major axis in the mm waveband, and possibly obtain the first polarization detections on VLBI baselines at 86\,GHz, allowing us to further sharpen our view of the accretion flow of \sgra.

\acknowledgements{We thank our anonymous referee for a thorough review and helpful suggestions, which improved the quality and clarity of the manuscript. We thank Vincent Fish for his helpful comments and careful review. This work is supported by the ERC Synergy Grant ``BlackHoleCam: Imaging the Event Horizon of Black Holes'', Grant 610058. We thank the National Science Foundation (AST-1716536, AST-1440254, AST-1935980, AST-2034306) and the Gordon and Betty Moore Foundation (GBMF-5278) for financial support of this work. This work was supported in part by the Black Hole Initiative, which is funded by grants from the John Templeton Foundation and the Gordon and Betty Moore Foundation to Harvard University. I.C. is supported by the National Research Foundation of Korea (NRF) via a Global PhD Fellowship Grant (NRF-2015H1A2A1033752). L.L. acknowledges the financial support of DGAPA, UNAM (projects IN112417 and IN112820), and CONACyT, M\'exico (projects 275201 -- Agencia Espacial M\'exicana; and 263356 -- Ciencia de Frontera). M.K. was supported by JSPS KAKENHI grant Nos. JP18K03656 and JP18H03721. 
This paper makes use of the following ALMA data: ADS/JAO.ALMA2017.1.00795.V. ALMA is a partnership of ESO (representing its member states), NSF (USA) and NINS (Japan), together with NRC (Canada), MOST and ASIAA (Taiwan), and KASI (Republic of Korea), in cooperation with the Republic of Chile. The Joint ALMA Observatory is operated by ESO, AUI/NRAO and NAOJ. The ALMA data required non-standard processing by the VLBI QA2 team (C. Goddi, I. Mart{\'i}-Vidal, G. B. Crew, H. Rottmann, and H. Messias). This research has made use of data obtained with the Global Millimeter VLBI Array (GMVA), coordinated by the VLBI group at the Max-Planck-Institut f{\"u}r Radioastronomie (MPIfR). The GMVA consists of telescopes operated by MPIfR, IRAM, Onsala, Metsahovi, Yebes, the Korean VLBI Network, the Green Bank Observatory and the Very Long Baseline Array (VLBA). The VLBA is a facility of the National Science Foundation under cooperative agreement by Associated Universities, Inc. The data were correlated at the DiFX correlator of the MPIfR in Bonn, Germany. This work made use of the Swinburne University of Technology software correlator \citep{Deller_2011}, developed as part of the Australian Major National Research Facilities Programme and operated under licence.}

\facilities{GMVA, ALMA, VLBA, GBT.}

\software{
AIPS \citep{Greisen_2003},
DiFX \citep{Deller_2011},
HOPS \citep{Whitney_2004,lindyhops},
PolConvert \citep{Marti_2016},
GNU Parallel \citep{GNU},
Numpy \citep{numpy_2011}, 
Scipy \citep{scipy}, 
Pandas \citep{pandas}, 
Astropy \citep{astropy_2013,astropy_2018}, 
Jupyter \citep{jupyter}, 
Matplotlib \citep{Hunter_2007},
eht-imaging \citep{Chael_2016},
\themis \citep{broderick20}
}

\bibliography{3mm_bib.bib}

\begin{thebibliography}{}
\expandafter\ifx\csname natexlab\endcsname\relax\def\natexlab#1{#1}\fi
\providecommand{\url}[1]{\href{#1}{#1}}

\bibitem[{{Alberdi} {et~al.}(1993){Alberdi}, {Lara}, {Marcaide}, {Elosegui},
  {Shapiro}, {Cotton}, {Diamond}, {Romney}, \& {Preston}}]{Alberdi_1993}
{Alberdi}, A., {Lara}, L., {Marcaide}, J.~M., {et~al.} 1993, \aap, 277, L1

\bibitem[{Backer {et~al.}(1993)Backer, Zensus, Kellermann, Reid, Moran, \&
  Lo}]{Backer_1993}
Backer, D.~C., Zensus, J.~A., Kellermann, K.~I., {et~al.} 1993, Science, 262,
  1414.
\newblock \url{https://science.sciencemag.org/content/262/5138/1414}

\bibitem[{{Balick} \& {Brown}(1974)}]{Balick_Brown_1974}
{Balick}, B., \& {Brown}, R.~L. 1974, \apj, 194, 265

\bibitem[{{Blackburn} {et~al.}(2020){Blackburn}, {Pesce}, {Johnson}, {Wielgus},
  {Chael}, {Christian}, \& {Doeleman}}]{lindyclosures}
{Blackburn}, L., {Pesce}, D.~W., {Johnson}, M.~D., {et~al.} 2020, \apj, 894, 31

\bibitem[{{Blackburn} {et~al.}(2019){Blackburn}, {Chan}, {Crew}, {Fish},
  {Issaoun}, {Johnson}, {Wielgus}, {Akiyama}, {Barrett}, {Bouman}, {Cappallo},
  {Chael}, {Janssen}, {Lonsdale}, \& {Doeleman}}]{lindyhops}
{Blackburn}, L., {Chan}, C.-k., {Crew}, G.~B., {et~al.} 2019, \apj, 882, 23

\bibitem[{{Blandford} \& {Narayan}(1985)}]{Blandford_Narayan_1985}
{Blandford}, R., \& {Narayan}, R. 1985, \mnras, 213, 591

\bibitem[{{Bower} \& {Backer}(1998)}]{Bower_1998}
{Bower}, G.~C., \& {Backer}, D.~C. 1998, \apjl, 507, L117

\bibitem[{{Bower} {et~al.}(1997){Bower}, {Backer}, {Wright}, {Forster},
  {Aller}, \& {Aller}}]{Bower_1997}
{Bower}, G.~C., {Backer}, D.~C., {Wright}, M., {et~al.} 1997, \apj, 484, 118

\bibitem[{{Bower} {et~al.}(2004){Bower}, {Falcke}, {Herrnstein}, {Zhao},
  {Goss}, \& {Backer}}]{Bower_2004}
{Bower}, G.~C., {Falcke}, H., {Herrnstein}, R.~M., {et~al.} 2004, Science, 304,
  704

\bibitem[{{Bower} {et~al.}(2006){Bower}, {Goss}, {Falcke}, {Backer}, \&
  {Lithwick}}]{Bower_2006}
{Bower}, G.~C., {Goss}, W.~M., {Falcke}, H., {Backer}, D.~C., \& {Lithwick}, Y.
  2006, \apjl, 648, L127

\bibitem[{{Bower} {et~al.}(2014){Bower}, {Markoff}, {Brunthaler}, {Law},
  {Falcke}, {Maitra}, {Clavel}, {Goldwurm}, {Morris}, {Witzel}, {Meyer}, \&
  {Ghez}}]{Bower_2014b}
{Bower}, G.~C., {Markoff}, S., {Brunthaler}, A., {et~al.} 2014, \apj, 790, 1

\bibitem[{{Bower} {et~al.}(2015){Bower}, {Deller}, {Demorest}, {Brunthaler},
  {Falcke}, {Moscibrodzka}, {O'Leary}, {Eatough}, {Kramer}, {Lee}, {Spitler},
  {Desvignes}, {Rushton}, {Doeleman}, \& {Reid}}]{Bower_2015}
{Bower}, G.~C., {Deller}, A., {Demorest}, P., {et~al.} 2015, \apj, 798, 120

\bibitem[{{Bower} {et~al.}(2019){Bower}, {Dexter}, {Asada}, {Brinkerink},
  {Falcke}, {Ho}, {Inoue}, {Markoff}, {Marrone}, {Matsushita}, {Moscibrodzka},
  {Nakamura}, {Peck}, \& {Rao}}]{Bower_2019}
{Bower}, G.~C., {Dexter}, J., {Asada}, K., {et~al.} 2019, \apjl, 881, L2

\bibitem[{{Brinkerink} {et~al.}(2016){Brinkerink}, {M{\"u}ller}, {Falcke},
  {Bower}, {Krichbaum}, {Castillo}, {Deller}, {Doeleman}, {Fraga-Encinas},
  {Goddi}, {Hern{\'a}ndez-G{\'o}mez}, {Hughes}, {Kramer}, {L{\'e}on-Tavares},
  {Loinard}, {Monta{\~n}a}, {Mo{\'s}cibrodzka}, {Ortiz-Le{\'o}n},
  {Sanchez-Arguelles}, {Tilanus}, {Wilson}, \& {Zensus}}]{Brinkerink_2016}
{Brinkerink}, C.~D., {M{\"u}ller}, C., {Falcke}, H., {et~al.} 2016, \mnras,
  462, 1382, (B16)

\bibitem[{{Brinkerink} {et~al.}(2019){Brinkerink}, {M{\"u}ller}, {Falcke},
  {Issaoun}, {Akiyama}, {Bower}, {Krichbaum}, {Deller}, {Castillo}, {Doeleman},
  {Fraga-Encinas}, {Goddi}, {Hern{\'a}ndez-G{\'o}mez}, {Hughes}, {Kramer},
  {L{\'e}on-Tavares}, {Loinard}, {Monta{\~n}a}, {Mo{\'s}cibrodzka},
  {Ortiz-Le{\'o}n}, {Sanchez-Arguelles}, {Tilanus}, {Wilson}, \&
  {Zensus}}]{Brinkerink_2019}
{Brinkerink}, C.~D., {M{\"u}ller}, C., {Falcke}, H.~D., {et~al.} 2019, \aap,
  621, A119, (B19)

\bibitem[{{Broderick} {et~al.}(2020){Broderick}, {Gold}, {Karami},
  {Preciado-L{\'o}pez}, {Tiede}, {Pu}, {Akiyama}, {Alberdi}, {Alef}, {Asada},
  {Azulay}, {Baczko}, {Balokovi{\'c}}, {Barrett}, {Bintley}, {Blackburn},
  {Boland}, {Bouman}, {Bower}, {Bremer}, {Brinkerink}, {Brissenden}, {Britzen},
  {Broguiere}, {Bronzwaer}, {Byun}, {Carlstrom}, {Chael}, {Chatterjee},
  {Chatterjee}, {Chen}, {Chen}, {Cho}, {Conway}, {Cordes}, {Crew}, {Cui},
  {Davelaar}, {De Laurentis}, {Deane}, {Dempsey}, {Desvignes}, {Doeleman},
  {Eatough}, {Falcke}, {Fish}, {Fomalont}, {Fraga-Encinas}, {Friberg}, {Fromm},
  {Galison}, {Gammie}, {Garc{\'\i}a}, {Gentaz}, {Georgiev}, {Goddi},
  {G{\'o}mez}, {Gu}, {Gurwell}, {Hada}, {Hecht}, {Hesper}, {Ho}, {Ho}, {Honma},
  {Huang}, {Huang}, {Hughes}, {Inoue}, {Issaoun}, {James}, {Janssen}, {Jeter},
  {Jiang}, {Jim{\'e}nez-Rosales}, {Johnson}, {Jorstad}, {Jung}, {Karuppusamy},
  {Kawashima}, {Keating}, {Kettenis}, {Kim}, {Kim}, {Kino}, {Koay}, {Koch},
  {Koyama}, {Kramer}, {Kramer}, {Krichbaum}, {Kuo}, {Lee}, {Li}, {Li},
  {Lindqvist}, {Lico}, {Liu}, {Liuzzo}, {Lo}, {Lobanov}, {Loinard}, {Lonsdale},
  {Lu}, {MacDonald}, {Mao}, {Marscher}, {Mart{\'\i}-Vidal}, {Matsushita},
  {Matthews}, {Menten}, {Mizuno}, {Mizuno}, {Moran}, {Moriyama},
  {Moscibrodzka}, {M{\"u}ller}, {Nagai}, {Nagar}, {Nakamura}, {Narayan},
  {Narayanan}, {Natarajan}, {Neri}, {Ni}, {Noutsos}, {Okino}, {Olivares},
  {Ortiz-Le{\'o}n}, {Oyama}, {Palumbo}, {Park}, {Pen}, {Pesce}, {Pi{\'e}tu},
  {Plambeck}, {PopStefanija}, {Porth}, {Prather}, {Ramakrishnan}, {Rao},
  {Rawlings}, {Raymond}, {Rezzolla}, {Ripperda}, {Roelofs}, {Rogers}, {Ros},
  {Rose}, {Rottmann}, {Ruszczyk}, {Ryan}, {Rygl}, {S{\'a}nchez},
  {S{\'a}nchez-Arguelles}, {Sasada}, {Savolainen}, {Schloerb}, {Schuster},
  {Shao}, {Shen}, {Small}, {Sohn}, {SooHoo}, {Tazaki}, {Tilanus}, {Titus},
  {Toma}, {Torne}, {Traianou}, {Trippe}, {Tsuda}, {van Bemmel}, {van
  Langevelde}, {van Rossum}, {Wagner}, {Wardle}, {Weintroub}, {Wex}, {Wharton},
  {Wielgus}, {Wong}, {Wu}, {Yoon}, {Young}, {Young}, {Younsi}, {Yuan}, {Yuan},
  {Zensus}, {Zhao}, {Zhao}, {Zhu}, \& {Event Horizon Telescope
  Collaboration}}]{broderick20}
{Broderick}, A.~E., {Gold}, R., {Karami}, M., {et~al.} 2020, \apj, 897, 139

\bibitem[{{Chael} {et~al.}(2018{\natexlab{a}}){Chael}, {Rowan}, {Narayan},
  {Johnson}, \& {Sironi}}]{Chael_2018b}
{Chael}, A., {Rowan}, M., {Narayan}, R., {Johnson}, M., \& {Sironi}, L.
  2018{\natexlab{a}}, \mnras, 478, 5209

\bibitem[{{Chael} {et~al.}(2018{\natexlab{b}}){Chael}, {Johnson}, {Bouman},
  {Blackburn}, {Akiyama}, \& {Narayan}}]{Chael_2018}
{Chael}, A.~A., {Johnson}, M.~D., {Bouman}, K.~L., {et~al.} 2018{\natexlab{b}},
  \apj, 857, 23

\bibitem[{{Chael} {et~al.}(2016){Chael}, {Johnson}, {Narayan}, {Doeleman},
  {Wardle}, \& {Bouman}}]{Chael_2016}
{Chael}, A.~A., {Johnson}, M.~D., {Narayan}, R., {et~al.} 2016, \apj, 829, 11

\bibitem[{{Chen} {et~al.}(2010){Chen}, {Shen}, \& {Feng}}]{Chen_2010}
{Chen}, Y.~J., {Shen}, Z.-Q., \& {Feng}, S.-W. 2010, \mnras, 408, 841

\bibitem[{Cornwell(1989)}]{Cornwell_1989}
Cornwell, T.~J. 1989, Science, 245, 263.
\newblock \url{https://science.sciencemag.org/content/245/4915/263}

\bibitem[{{Davelaar} {et~al.}(2018){Davelaar}, {Mo{\'s}cibrodzka}, {Bronzwaer},
  \& {Falcke}}]{Davelaar_2018}
{Davelaar}, J., {Mo{\'s}cibrodzka}, M., {Bronzwaer}, T., \& {Falcke}, H. 2018,
  \aap, 612, A34

\bibitem[{{Davies} {et~al.}(1976){Davies}, {Walsh}, \& {Booth}}]{Davies_1976}
{Davies}, R.~D., {Walsh}, D., \& {Booth}, R.~S. 1976, \mnras, 177, 319

\bibitem[{{Deller} {et~al.}(2011){Deller}, {Brisken}, {Phillips}, {Morgan},
  {Alef}, {Cappallo}, {Middelberg}, {Romney}, {Rottmann}, {Tingay}, \&
  {Wayth}}]{Deller_2011}
{Deller}, A.~T., {Brisken}, W.~F., {Phillips}, C.~J., {et~al.} 2011, \pasp,
  123, 275

\bibitem[{{Doeleman} {et~al.}(2001){Doeleman}, {Shen}, {Rogers}, {Bower},
  {Wright}, {Zhao}, {Backer}, {Crowley}, {Freund}, {Ho}, {Lo}, \&
  {Woody}}]{Doeleman_2001}
{Doeleman}, S.~S., {Shen}, Z.-Q., {Rogers}, A.~E.~E., {et~al.} 2001, \aj, 121,
  2610

\bibitem[{{Doeleman} {et~al.}(2008){Doeleman}, {Weintroub}, {Rogers},
  {Plambeck}, {Freund}, {Tilanus}, {Friberg}, {Ziurys}, {Moran}, {Corey},
  {Young}, {Smythe}, {Titus}, {Marrone}, {Cappallo}, {Bock}, {Bower},
  {Chamberlin}, {Davis}, {Krichbaum}, {Lamb}, {Maness}, {Niell}, {Roy},
  {Strittmatter}, {Werthimer}, {Whitney}, \& {Woody}}]{Doeleman_2008}
{Doeleman}, S.~S., {Weintroub}, J., {Rogers}, A.~E.~E., {et~al.} 2008, \nat,
  455, 78

\bibitem[{{Event Horizon Telescope Collaboration} {et~al.}(2019){Event Horizon
  Telescope Collaboration}, {Akiyama}, {Alberdi}, {Alef}, {Asada}, {Azulay},
  {Baczko}, {Ball}, {Balokovi{\'c}}, {Barrett}, \& et~al.}]{PaperI}
{Event Horizon Telescope Collaboration}, {Akiyama}, K., {Alberdi}, A., {et~al.}
  2019, \apjl, 875, L1

\bibitem[{{Falcke} {et~al.}(1998){Falcke}, {Goss}, {Matsuo}, {Teuben}, {Zhao},
  \& {Zylka}}]{Falcke_1998}
{Falcke}, H., {Goss}, W.~M., {Matsuo}, H., {et~al.} 1998, \apj, 499, 731

\bibitem[{{Falcke} \& {Markoff}(2000)}]{Falcke_Markoff_2000}
{Falcke}, H., \& {Markoff}, S. 2000, \aap, 362, 113

\bibitem[{{Feng} {et~al.}(2006){Feng}, {Shen}, {Cai}, {Chen}, {Lu}, \&
  {Huang}}]{Feng_2006}
{Feng}, S.-W., {Shen}, Z.-Q., {Cai}, H.-B., {et~al.} 2006, \aap, 456, 97

\bibitem[{{Fish} {et~al.}(2011){Fish}, {Doeleman}, {Beaudoin}, {Blundell},
  {Bolin}, {Bower}, {Chamberlin}, {Freund}, {Friberg}, {Gurwell}, {Honma},
  {Inoue}, {Krichbaum}, {Lamb}, {Marrone}, {Moran}, {Oyama}, {Plambeck},
  {Primiani}, {Rogers}, {Smythe}, {SooHoo}, {Strittmatter}, {Tilanus}, {Titus},
  {Weintroub}, {Wright}, {Woody}, {Young}, \& {Ziurys}}]{Fish_2011}
{Fish}, V.~L., {Doeleman}, S.~S., {Beaudoin}, C., {et~al.} 2011, \apjl, 727,
  L36

\bibitem[{{Fish} {et~al.}(2016){Fish}, {Johnson}, {Doeleman}, {Broderick},
  {Psaltis}, {Lu}, {Akiyama}, {Alef}, {Algaba}, {Asada}, {Beaudoin},
  {Bertarini}, {Blackburn}, {Blundell}, {Bower}, {Brinkerink}, {Cappallo},
  {Chael}, {Chamberlin}, {Chan}, {Crew}, {Dexter}, {Dexter}, {Dzib}, {Falcke},
  {Freund}, {Friberg}, {Greer}, {Gurwell}, {Ho}, {Honma}, {Inoue}, {Johannsen},
  {Kim}, {Krichbaum}, {Lamb}, {Le{\'o}n-Tavares}, {Loeb}, {Loinard},
  {MacMahon}, {Marrone}, {Moran}, {Mo{\'s}cibrodzka}, {Ortiz-Le{\'o}n},
  {Oyama}, {{\"O}zel}, {Plambeck}, {Pradel}, {Primiani}, {Rogers}, {Rosenfeld},
  {Rottmann}, {Roy}, {Ruszczyk}, {Smythe}, {SooHoo}, {Spilker}, {Stone},
  {Strittmatter}, {Tilanus}, {Titus}, {Vertatschitsch}, {Wagner}, {Wardle},
  {Weintroub}, {Woody}, {Wright}, {Yamaguchi}, {Young}, {Young}, {Zensus}, \&
  {Ziurys}}]{Fish_2016}
{Fish}, V.~L., {Johnson}, M.~D., {Doeleman}, S.~S., {et~al.} 2016, \apj, 820,
  90

\bibitem[{{Frail} {et~al.}(1994){Frail}, {Diamond}, {Cordes}, \& {van
  Langevelde}}]{Frail_1994}
{Frail}, D.~A., {Diamond}, P.~J., {Cordes}, J.~M., \& {van Langevelde}, H.~J.
  1994, \apj, 427, L43

\bibitem[{{Ghez} {et~al.}(2008){Ghez}, {Salim}, {Weinberg}, {Lu}, {Do}, {Dunn},
  {Matthews}, {Morris}, {Yelda}, {Becklin}, {Kremenek}, {Milosavljevic}, \&
  {Naiman}}]{Ghez_2008}
{Ghez}, A.~M., {Salim}, S., {Weinberg}, N.~N., {et~al.} 2008, \apj, 689, 1044

\bibitem[{{Gillessen} {et~al.}(2009){Gillessen}, {Eisenhauer}, {Fritz},
  {Bartko}, {Dodds-Eden}, {Pfuhl}, {Ott}, \& {Genzel}}]{Gillessen_2009}
{Gillessen}, S., {Eisenhauer}, F., {Fritz}, T.~K., {et~al.} 2009, \apjl, 707,
  L114

\bibitem[{{Goddi} {et~al.}(2017){Goddi}, {Falcke}, {Kramer}, {Rezzolla},
  {Brinkerink}, {Bronzwaer}, {Davelaar}, {Deane}, {de Laurentis}, {Desvignes},
  {Eatough}, {Eisenhauer}, {Fraga-Encinas}, {Fromm}, {Gillessen}, {Grenzebach},
  {Issaoun}, {Jan{\ss}en}, {Konoplya}, {Krichbaum}, {Laing}, {Liu}, {Lu},
  {Mizuno}, {Moscibrodzka}, {M{\"u}ller}, {Olivares}, {Pfuhl}, {Porth},
  {Roelofs}, {Ros}, {Schuster}, {Tilanus}, {Torne}, {van Bemmel}, {van
  Langevelde}, {Wex}, {Younsi}, \& {Zhidenko}}]{Goddi_2016}
{Goddi}, C., {Falcke}, H., {Kramer}, M., {et~al.} 2017, International Journal
  of Modern Physics D, 26, 1730001

\bibitem[{{Goddi} {et~al.}(2019){Goddi}, {Mart{\'\i}-Vidal}, {Messias}, {Crew},
  {Herrero-Illana}, {Impellizzeri}, {Rottmann}, {Wagner}, {Fomalont},
  {Matthews}, {Petry}, {Phillips}, {Tilanus}, {Villard}, {Blackburn},
  {Janssen}, \& {Wielgus}}]{Goddi_2018}
{Goddi}, C., {Mart{\'\i}-Vidal}, I., {Messias}, H., {et~al.} 2019, \pasp, 131,
  075003

\bibitem[{{Goldreich} \& {Sridhar}(2006)}]{Goldreich_2006}
{Goldreich}, P., \& {Sridhar}, S. 2006, \apjl, 640, L159, (GS06)

\bibitem[{{Goodman} \& {Narayan}(1989)}]{GoodmanNarayan89}
{Goodman}, J., \& {Narayan}, R. 1989, \mnras, 238, 995

\bibitem[{{Gravity Collaboration} {et~al.}(2018{\natexlab{a}}){Gravity
  Collaboration}, {Abuter}, {Amorim}, {Anugu}, {Baub{\"o}ck}, {Benisty},
  {Berger}, {Blind}, {Bonnet}, {Brandner}, {Buron}, {Collin}, {Chapron},
  {Cl{\'e}net}, {Coud{\'e} Du Foresto}, {de Zeeuw}, {Deen},
  {Delplancke-Str{\"o}bele}, {Dembet}, {Dexter}, {Duvert}, {Eckart},
  {Eisenhauer}, {Finger}, {F{\"o}rster Schreiber}, {F{\'e}dou}, {Garcia},
  {Garcia Lopez}, {Gao}, {Gendron}, {Genzel}, {Gillessen}, {Gordo}, {Habibi},
  {Haubois}, {Haug}, {Hau{\ss}mann}, {Henning}, {Hippler}, {Horrobin},
  {Hubert}, {Hubin}, {Jimenez Rosales}, {Jochum}, {Jocou}, {Kaufer}, {Kellner},
  {Kendrew}, {Kervella}, {Kok}, {Kulas}, {Lacour}, {Lapeyr{\`e}re}, {Lazareff},
  {Le Bouquin}, {L{\'e}na}, {Lippa}, {Lenzen}, {M{\'e}rand}, {M{\"u}ler},
  {Neumann}, {Ott}, {Palanca}, {Paumard}, {Pasquini}, {Perraut}, {Perrin},
  {Pfuhl}, {Plewa}, {Rabien}, {Ram{\'{\i}}rez}, {Ramos}, {Rau},
  {Rodr{\'{\i}}guez-Coira}, {Rohloff}, {Rousset}, {Sanchez-Bermudez},
  {Scheithauer}, {Sch{\"o}ller}, {Schuler}, {Spyromilio}, {Straub},
  {Straubmeier}, {Sturm}, {Tacconi}, {Tristram}, {Vincent}, {von Fellenberg},
  {Wank}, {Waisberg}, {Widmann}, {Wieprecht}, {Wiest}, {Wiezorrek}, {Woillez},
  {Yazici}, {Ziegler}, \& {Zins}}]{Gravity_2018}
{Gravity Collaboration}, {Abuter}, R., {Amorim}, A., {et~al.}
  2018{\natexlab{a}}, \aap, 615, L15

\bibitem[{{Gravity Collaboration} {et~al.}(2018{\natexlab{b}}){Gravity
  Collaboration}, {Abuter}, {Amorim}, {Baub{\"o}ck}, {Berger}, {Bonnet},
  {Brandner}, {Cl{\'e}net}, {Coud{\'e} Du Foresto}, {de Zeeuw}, {Deen},
  {Dexter}, {Duvert}, {Eckart}, {Eisenhauer}, {F{\"o}rster Schreiber},
  {Garcia}, {Gao}, {Gendron}, {Genzel}, {Gillessen}, {Guajardo}, {Habibi},
  {Haubois}, {Henning}, {Hippler}, {Horrobin}, {Huber}, {Jim{\'e}nez-Rosales},
  {Jocou}, {Kervella}, {Lacour}, {Lapeyr{\`e}re}, {Lazareff}, {Le Bouquin},
  {L{\'e}na}, {Lippa}, {Ott}, {Panduro}, {Paumard}, {Perraut}, {Perrin},
  {Pfuhl}, {Plewa}, {Rabien}, {Rodr{\'{\i}}guez-Coira}, {Rousset}, {Sternberg},
  {Straub}, {Straubmeier}, {Sturm}, {Tacconi}, {Vincent}, {von Fellenberg},
  {Waisberg}, {Widmann}, {Wieprecht}, {Wiezorrek}, {Woillez}, \&
  {Yazici}}]{Gravity_2018b}
---. 2018{\natexlab{b}}, \aap, 618, L10

\bibitem[{{Greisen}(2003)}]{Greisen_2003}
{Greisen}, E.~W. 2003, in Astrophysics and Space Science Library, Vol. 285,
  Information Handling in Astronomy - Historical Vistas, ed. A.~{Heck} (Kluwer
  Academic Publishers), 109

\bibitem[{{Gwinn} {et~al.}(2014){Gwinn}, {Kovalev}, {Johnson}, \&
  {Soglasnov}}]{Gwinn_2014}
{Gwinn}, C.~R., {Kovalev}, Y.~Y., {Johnson}, M.~D., \& {Soglasnov}, V.~A. 2014,
  \apjl, 794, L14

\bibitem[{{Howes}(2010)}]{Howes_2010}
{Howes}, G.~G. 2010, \mnras, 409, L104

\bibitem[{Hu(1962)}]{Hu_1962}
Hu, M.-K. 1962, Information Theory, IRE Transactions on, 8, 179

\bibitem[{Hunter(2007)}]{Hunter_2007}
Hunter, J.~D. 2007, Computing In Science \& Engineering, 9, 90

\bibitem[{{Issaoun} {et~al.}(2019{\natexlab{a}}){Issaoun}, {Johnson},
  {Blackburn}, {Mo{\'s}cibrodzka}, {Chael}, \& {Falcke}}]{Issaoun_2019b}
{Issaoun}, S., {Johnson}, M.~D., {Blackburn}, L., {et~al.} 2019{\natexlab{a}},
  \aap, 629, A32

\bibitem[{{Issaoun} {et~al.}(2019{\natexlab{b}}){Issaoun}, {Johnson},
  {Blackburn}, {Brinkerink}, {Mo{\'s}cibrodzka}, {Chael}, {Goddi},
  {Mart{\'{\i}}-Vidal}, {Wagner}, {Doeleman}, {Falcke}, {Krichbaum}, {Akiyama},
  {Bach}, {Bouman}, {Bower}, {Broderick}, {Cho}, {Crew}, {Dexter}, {Fish},
  {Gold}, {G{\'o}mez}, {Hada}, {Hern{\'a}ndez-G{\'o}mez}, {Jan{\ss}en}, {Kino},
  {Kramer}, {Loinard}, {Lu}, {Markoff}, {Marrone}, {Matthews}, {Moran},
  {M{\"u}ller}, {Roelofs}, {Ros}, {Rottmann}, {Sanchez}, {Tilanus}, {de
  Vicente}, {Wielgus}, {Zensus}, \& {Zhao}}]{Issaoun_2019}
---. 2019{\natexlab{b}}, \apj, 871, 30

\bibitem[{{Johnson}(2016)}]{Johnson_2016}
{Johnson}, M.~D. 2016, \apj, 833, 74

\bibitem[{{Johnson} \& {Gwinn}(2015)}]{Johnson_Gwinn_2015}
{Johnson}, M.~D., \& {Gwinn}, C.~R. 2015, \apj, 805, 180

\bibitem[{{Johnson} \& {Narayan}(2016)}]{Johnson_Narayan_2016}
{Johnson}, M.~D., \& {Narayan}, R. 2016, \apj, 826, 170

\bibitem[{{Johnson} {et~al.}(2015){Johnson}, {Fish}, {Doeleman}, {Marrone},
  {Plambeck}, {Wardle}, {Akiyama}, {Asada}, {Beaudoin}, {Blackburn},
  {Blundell}, {Bower}, {Brinkerink}, {Broderick}, {Cappallo}, {Chael}, {Crew},
  {Dexter}, {Dexter}, {Freund}, {Friberg}, {Gold}, {Gurwell}, {Ho}, {Honma},
  {Inoue}, {Kosowsky}, {Krichbaum}, {Lamb}, {Loeb}, {Lu}, {MacMahon},
  {McKinney}, {Moran}, {Narayan}, {Primiani}, {Psaltis}, {Rogers}, {Rosenfeld},
  {SooHoo}, {Tilanus}, {Titus}, {Vertatschitsch}, {Weintroub}, {Wright},
  {Young}, {Zensus}, \& {Ziurys}}]{Johnson_2015}
{Johnson}, M.~D., {Fish}, V.~L., {Doeleman}, S.~S., {et~al.} 2015, Science,
  350, 1242

\bibitem[{{Johnson} {et~al.}(2018){Johnson}, {Narayan}, {Psaltis}, {Blackburn},
  {Kovalev}, {Gwinn}, {Zhao}, {Bower}, {Moran}, {Kino}, {Kramer}, {Akiyama},
  {Dexter}, {Broderick}, \& {Sironi}}]{Johnson_2018}
{Johnson}, M.~D., {Narayan}, R., {Psaltis}, D., {et~al.} 2018, \apj, 865, 104,
  (J18)

\bibitem[{Jones {et~al.}(2001)Jones, Oliphant, Peterson, {et~al.}}]{scipy}
Jones, E., Oliphant, T., Peterson, P., {et~al.} 2001, {SciPy}: Open source
  scientific tools for {Python}, , .
\newblock \url{http://www.scipy.org/}

\bibitem[{Kluyver {et~al.}(2016)Kluyver, Ragan-Kelley, P{\'e}rez, Granger,
  Bussonnier, Frederic, Kelley, Hamrick, Grout, Corlay, Ivanov, Avila, Abdalla,
  \& Willing}]{jupyter}
Kluyver, T., Ragan-Kelley, B., P{\'e}rez, F., {et~al.} 2016, in Positioning and
  Power in Academic Publishing: Players, Agents and Agendas, ed. F.~Loizides \&
  B.~Schmidt (IOS Press), 87 -- 90

\bibitem[{{Krichbaum} {et~al.}(1993){Krichbaum}, {Zensus}, {Witzel}, {Mezger},
  {Standke}, {Schalinski}, {Alberdi}, {Marcaide}, {Zylka}, {Rogers}, {Booth},
  {Ronnang}, {Colomer}, {Bartel}, \& {Shapiro}}]{Krichbaum_1993}
{Krichbaum}, T.~P., {Zensus}, J.~A., {Witzel}, A., {et~al.} 1993, \aap, 274,
  L37

\bibitem[{{Krichbaum} {et~al.}(1998){Krichbaum}, {Graham}, {Witzel}, {Greve},
  {Wink}, {Grewing}, {Colomer}, {de Vicente}, {Gomez-Gonzalez}, {Baudry}, \&
  {Zensus}}]{Krichbaum_1998}
{Krichbaum}, T.~P., {Graham}, D.~A., {Witzel}, A., {et~al.} 1998, \aap, 335,
  L106

\bibitem[{{Lu} {et~al.}(2011{\natexlab{a}}){Lu}, {Krichbaum}, {Eckart},
  {K{\"o}nig}, {Kunneriath}, {Witzel}, {Witzel}, \& {Zensus}}]{Lu_2011}
{Lu}, R.-S., {Krichbaum}, T.~P., {Eckart}, A., {et~al.} 2011{\natexlab{a}},
  \aap, 525, A76

\bibitem[{{Lu} {et~al.}(2011{\natexlab{b}}){Lu}, {Krichbaum}, \&
  {Zensus}}]{Lu_2011b}
{Lu}, R.-S., {Krichbaum}, T.~P., \& {Zensus}, J.~A. 2011{\natexlab{b}}, \mnras,
  418, 2260

\bibitem[{{Lu} {et~al.}(2012){Lu}, {Fish}, {Weintroub}, {Doeleman}, {Bower},
  {Freund}, {Friberg}, {Ho}, {Honma}, {Inoue}, {Krichbaum}, {Marrone}, {Moran},
  {Oyama}, {Plambeck}, {Primiani}, {Shen}, {Tilanus}, {Wright}, {Young},
  {Ziurys}, \& {Zensus}}]{Lu_2012}
{Lu}, R.-S., {Fish}, V.~L., {Weintroub}, J., {et~al.} 2012, \apjl, 757, L14

\bibitem[{{Lu} {et~al.}(2018){Lu}, {Krichbaum}, {Roy}, {Fish}, {Doeleman},
  {Johnson}, {Akiyama}, {Psaltis}, {Alef}, {Asada}, {Beaudoin}, {Bertarini},
  {Blackburn}, {Blundell}, {Bower}, {Brinkerink}, {Broderick}, {Cappallo},
  {Crew}, {Dexter}, {Dexter}, {Falcke}, {Freund}, {Friberg}, {Greer},
  {Gurwell}, {Ho}, {Honma}, {Inoue}, {Kim}, {Lamb}, {Lindqvist}, {Macmahon},
  {Marrone}, {Mart{\'{\i}}-Vidal}, {Menten}, {Moran}, {Nagar}, {Plambeck},
  {Primiani}, {Rogers}, {Ros}, {Rottmann}, {SooHoo}, {Spilker}, {Stone},
  {Strittmatter}, {Tilanus}, {Titus}, {Vertatschitsch}, {Wagner}, {Weintroub},
  {Wright}, {Young}, {Zensus}, \& {Ziurys}}]{Lu_2018}
{Lu}, R.-S., {Krichbaum}, T.~P., {Roy}, A.~L., {et~al.} 2018, \apj, 859, 60

\bibitem[{{Marcaide} {et~al.}(1999){Marcaide}, {Alberdi}, {Lara},
  {P{\'e}rez-Torres}, \& {Diamond}}]{Marcaide_1999}
{Marcaide}, J.~M., {Alberdi}, A., {Lara}, L., {P{\'e}rez-Torres}, M.~A., \&
  {Diamond}, P.~J. 1999, \aap, 343, 801

\bibitem[{{Mart{\'{\i}}-Vidal} {et~al.}(2016){Mart{\'{\i}}-Vidal}, {Vlemmings},
  \& {Muller}}]{Marti_2016}
{Mart{\'{\i}}-Vidal}, I., {Vlemmings}, W.~H.~T., \& {Muller}, S. 2016, \aap,
  593, A61

\bibitem[{{Matthews} {et~al.}(2018){Matthews}, {Crew}, {Doeleman}, {Lacasse},
  {Saez}, {Alef}, {Akiyama}, {Amestica}, {Anderson}, {Barkats}, {Baudry},
  {Brogui{\`e}re}, {Escoffier}, {Fish}, {Greenberg}, {Hecht}, {Hiriart},
  {Hirota}, {Honma}, {Ho}, {Impellizzeri}, {Inoue}, {Kohno}, {Lopez},
  {Mart{\'{\i}}-Vidal}, {Messias}, {Meyer-Zhao}, {Mora-Klein}, {Nagar},
  {Nishioka}, {Oyama}, {Pankratius}, {Perez}, {Phillips}, {Pradel}, {Rottmann},
  {Roy}, {Ruszczyk}, {Shillue}, {Suzuki}, \& {Treacy}}]{Matthews_2018}
{Matthews}, L.~D., {Crew}, G.~B., {Doeleman}, S.~S., {et~al.} 2018, \pasp, 130,
  015002

\bibitem[{McKinney(2010)}]{pandas}
McKinney, W. 2010, in Proceedings of the 9th Python in Science Conference, ed.
  S.~van~der Walt \& J.~Millman, 51 -- 56

\bibitem[{{Mo{\'s}cibrodzka} {et~al.}(2016){Mo{\'s}cibrodzka}, {Falcke}, \&
  {Noble}}]{Moscibrodzka_2016}
{Mo{\'s}cibrodzka}, M., {Falcke}, H., \& {Noble}, S. 2016, \aap, 596, A13

\bibitem[{{Mo{\'s}cibrodzka} {et~al.}(2014){Mo{\'s}cibrodzka}, {Falcke},
  {Shiokawa}, \& {Gammie}}]{Moscibrodzka_2014}
{Mo{\'s}cibrodzka}, M., {Falcke}, H., {Shiokawa}, H., \& {Gammie}, C.~F. 2014,
  \aap, 570, A7

\bibitem[{{Mo{\'s}cibrodzka} {et~al.}(2009){Mo{\'s}cibrodzka}, {Gammie},
  {Dolence}, {Shiokawa}, \& {Leung}}]{Moscibrodzka_2009}
{Mo{\'s}cibrodzka}, M., {Gammie}, C.~F., {Dolence}, J.~C., {Shiokawa}, H., \&
  {Leung}, P.~K. 2009, \apj, 706, 497

\bibitem[{{Narayan} \& {Goodman}(1989)}]{NarayanGoodman89}
{Narayan}, R., \& {Goodman}, J. 1989, \mnras, 238, 963

\bibitem[{{Narayan} {et~al.}(1995){Narayan}, {Yi}, \&
  {Mahadevan}}]{Narayan_1995}
{Narayan}, R., {Yi}, I., \& {Mahadevan}, R. 1995, \nat, 374, 623

\bibitem[{{Ortiz-Le{\'o}n} {et~al.}(2016){Ortiz-Le{\'o}n}, {Johnson},
  {Doeleman}, {Blackburn}, {Fish}, {Loinard}, {Reid}, {Castillo}, {Chael},
  {Hern{\'a}ndez-G{\'o}mez}, {Hughes}, {Le{\'o}n-Tavares}, {Lu}, {Monta{\~n}a},
  {Narayanan}, {Rosenfeld}, {S{\'a}nchez}, {Schloerb}, {Shen}, {Shiokawa},
  {SooHoo}, \& {Vertatschitsch}}]{Ortiz_2016}
{Ortiz-Le{\'o}n}, G.~N., {Johnson}, M.~D., {Doeleman}, S.~S., {et~al.} 2016,
  \apj, 824, 40, (O16)

\bibitem[{{{\"O}zel} {et~al.}(2000){{\"O}zel}, {Psaltis}, \&
  {Narayan}}]{Ozel_2000}
{{\"O}zel}, F., {Psaltis}, D., \& {Narayan}, R. 2000, \apj, 541, 234

\bibitem[{{Psaltis} {et~al.}(2018){Psaltis}, {Johnson}, {Narayan}, {Medeiros},
  {Blackburn}, \& {Bower}}]{Psaltis_2018}
{Psaltis}, D., {Johnson}, M., {Narayan}, R., {et~al.} 2018, ArXiv e-prints,
  arXiv:1805.01242

\bibitem[{{Rauch} {et~al.}(2016){Rauch}, {Ros}, {Krichbaum}, {Eckart},
  {Zensus}, {Shahzamanian}, \& {Muzi\'{}c}}]{Rauch_2016}
{Rauch}, C., {Ros}, E., {Krichbaum}, T.~P., {et~al.} 2016, A\&A, 587, A37

\bibitem[{{Readhead} {et~al.}(1980){Readhead}, {Walker}, {Pearson}, \&
  {Cohen}}]{Readhead1980}
{Readhead}, A.~C.~S., {Walker}, R.~C., {Pearson}, T.~J., \& {Cohen}, M.~H.
  1980, \nat, 285, 137

\bibitem[{{Reid}(2009)}]{Reid_2009}
{Reid}, M.~J. 2009, International Journal of Modern Physics D, 18, 889

\bibitem[{{Rickett}(1990)}]{Rickett_1990}
{Rickett}, B.~J. 1990, \araa, 28, 561

\bibitem[{{Rogers} {et~al.}(1994){Rogers}, {Doeleman}, {Wright}, {Bower},
  {Backer}, {Padin}, {Philips}, {Emerson}, {Greenhill}, {Moran}, \&
  {Kellermann}}]{Rogers_1994}
{Rogers}, A.~E.~E., {Doeleman}, S., {Wright}, M.~C.~H., {et~al.} 1994, \apjl,
  434, L59

\bibitem[{{Rowan} {et~al.}(2017){Rowan}, {Sironi}, \& {Narayan}}]{Rowan_2017}
{Rowan}, M.~E., {Sironi}, L., \& {Narayan}, R. 2017, \apj, 850, 29

\bibitem[{{Shen} {et~al.}(2005){Shen}, {Lo}, {Liang}, {Ho}, \&
  {Zhao}}]{Shen_2005}
{Shen}, Z.-Q., {Lo}, K.~Y., {Liang}, M.-C., {Ho}, P.~T.~P., \& {Zhao}, J.-H.
  2005, \nat, 438, 62

\bibitem[{{Shen} {et~al.}(2002){Shen}, {Moran}, \& {Kellermann}}]{Shen_2002}
{Shen}, Z.-Q., {Moran}, J.~M., \& {Kellermann}, K.~I. 2002, in 8th
  Asian-Pacific Regional Meeting, Volume II, ed. S.~{Ikeuchi}, J.~{Hearnshaw},
  \& T.~{Hanawa} (Hitotsubashi Memorial Hall, Tokyo, Japan: The Astronomical
  Society of Japan), 401--402

\bibitem[{{Spangler} \& {Gwinn}(1990)}]{Spangler_Gwinn_1990}
{Spangler}, S.~R., \& {Gwinn}, C.~R. 1990, \apjl, 353, L29

\bibitem[{{Steel} {et~al.}(2019){Steel}, {Wielgus}, {Blackburn}, {Issaoun}, \&
  {Johnson}}]{Steel2019}
{Steel}, S., {Wielgus}, M., {Blackburn}, L., {Issaoun}, S., \& {Johnson}, M.
  2019, EHT Memo Series, 2019-CE-03

\bibitem[{{Syed} {et~al.}(2019){Syed}, {Bouchard-C{\^o}t{\'e}},
  {Deligiannidis}, \& {Doucet}}]{DEO:2019}
{Syed}, S., {Bouchard-C{\^o}t{\'e}}, A., {Deligiannidis}, G., \& {Doucet}, A.
  2019, arXiv e-prints, arXiv:1905.02939

\bibitem[{Tange(2011)}]{GNU}
Tange, O. 2011, ;login: The USENIX Magazine, 36, 42.
\newblock \url{http://www.gnu.org/s/parallel}

\bibitem[{{The Astropy Collaboration} {et~al.}(2013){The Astropy
  Collaboration}, {Robitaille}, {Tollerud}, {Greenfield}, {Droettboom}, {Bray},
  {Aldcroft}, {Davis}, {Ginsburg}, {Price-Whelan}, {Kerzendorf}, {Conley},
  {Crighton}, {Barbary}, {Muna}, {Ferguson}, {Grollier}, {Parikh}, {Nair},
  {Unther}, {Deil}, {Woillez}, {Conseil}, {Kramer}, {Turner}, {Singer}, {Fox},
  {Weaver}, {Zabalza}, {Edwards}, {Azalee Bostroem}, {Burke}, {Casey},
  {Crawford}, {Dencheva}, {Ely}, {Jenness}, {Labrie}, {Lim}, {Pierfederici},
  {Pontzen}, {Ptak}, {Refsdal}, {Servillat}, \& {Streicher}}]{astropy_2013}
{The Astropy Collaboration}, {Robitaille}, T.~P., {Tollerud}, E.~J., {et~al.}
  2013, \aap, 558, A33

\bibitem[{{The Astropy Collaboration} {et~al.}(2018){The Astropy
  Collaboration}, {Price-Whelan}, {Sip{\H o}cz}, {G{\"u}nther}, {Lim},
  {Crawford}, {Conseil}, {Shupe}, {Craig}, {Dencheva}, {Ginsburg},
  {VanderPlas}, {Bradley}, {P{\'e}rez-Su{\'a}rez}, {de Val-Borro}, {Aldcroft},
  {Cruz}, {Robitaille}, {Tollerud}, {Ardelean}, {Babej}, {Bach}, {Bachetti},
  {Bakanov}, {Bamford}, {Barentsen}, {Barmby}, {Baumbach}, {Berry}, {Biscani},
  {Boquien}, {Bostroem}, {Bouma}, {Brammer}, {Bray}, {Breytenbach},
  {Buddelmeijer}, {Burke}, {Calderone}, {Cano Rodr{\'{\i}}guez}, {Cara},
  {Cardoso}, {Cheedella}, {Copin}, {Corrales}, {Crichton}, {D'Avella}, {Deil},
  {Depagne}, {Dietrich}, {Donath}, {Droettboom}, {Earl}, {Erben}, {Fabbro},
  {Ferreira}, {Finethy}, {Fox}, {Garrison}, {Gibbons}, {Goldstein}, {Gommers},
  {Greco}, {Greenfield}, {Groener}, {Grollier}, {Hagen}, {Hirst}, {Homeier},
  {Horton}, {Hosseinzadeh}, {Hu}, {Hunkeler}, {Ivezi{\'c}}, {Jain}, {Jenness},
  {Kanarek}, {Kendrew}, {Kern}, {Kerzendorf}, {Khvalko}, {King}, {Kirkby},
  {Kulkarni}, {Kumar}, {Lee}, {Lenz}, {Littlefair}, {Ma}, {Macleod},
  {Mastropietro}, {McCully}, {Montagnac}, {Morris}, {Mueller}, {Mumford},
  {Muna}, {Murphy}, {Nelson}, {Nguyen}, {Ninan}, {N{\"o}the}, {Ogaz}, {Oh},
  {Parejko}, {Parley}, {Pascual}, {Patil}, {Patil}, {Plunkett}, {Prochaska},
  {Rastogi}, {Reddy Janga}, {Sabater}, {Sakurikar}, {Seifert}, {Sherbert},
  {Sherwood-Taylor}, {Shih}, {Sick}, {Silbiger}, {Singanamalla}, {Singer},
  {Sladen}, {Sooley}, {Sornarajah}, {Streicher}, {Teuben}, {Thomas},
  {Tremblay}, {Turner}, {Terr{\'o}n}, {van Kerkwijk}, {de la Vega}, {Watkins},
  {Weaver}, {Whitmore}, {Woillez}, {Zabalza}, \& {Astropy
  Contributors}}]{astropy_2018}
{The Astropy Collaboration}, {Price-Whelan}, A.~M., {Sip{\H o}cz}, B.~M.,
  {et~al.} 2018, \aj, 156, 123

\bibitem[{{Thompson} {et~al.}(2017){Thompson}, {Moran}, \& {Swenson}}]{TMS}
{Thompson}, A.~R., {Moran}, J.~M., \& {Swenson}, Jr., G.~W. 2017,
  {Interferometry and Synthesis in Radio Astronomy, 3rd Edition} ("Springer
  International Publishing"), doi:10.1007/978-3-319-44431-4

\bibitem[{{Tibbits} {et~al.}(2014){Tibbits}, {Groendyke}, {Hara}, \&
  C.}]{AFSS:2014}
{Tibbits}, M.~M., {Groendyke}, C., {Hara}, M., \& C., L.~J. 2014, J Comput
  Graph Stat., 23, 543

\bibitem[{{Twiss} {et~al.}(1960){Twiss}, {Carter}, \& {Little}}]{Twiss_1960}
{Twiss}, R.~Q., {Carter}, A.~W.~L., \& {Little}, A.~G. 1960, The Observatory,
  80, 153.
\newblock \url{https://ui.adsabs.harvard.edu/abs/1960Obs....80..153T}

\bibitem[{{van der Walt} {et~al.}(2011){van der Walt}, {Colbert}, \&
  {Varoquaux}}]{numpy_2011}
{van der Walt}, S., {Colbert}, S.~C., \& {Varoquaux}, G. 2011, Computing in
  Science and Engineering, 13, 22

\bibitem[{{van Langevelde} {et~al.}(1992){van Langevelde}, {Frail}, {Cordes},
  \& {Diamond}}]{vanLangevelde_1992}
{van Langevelde}, H.~J., {Frail}, D.~A., {Cordes}, J.~M., \& {Diamond}, P.~J.
  1992, \apj, 396, 686

\bibitem[{Whitney {et~al.}(2004)Whitney, Cappallo, Aldrich, Anderson, Bos,
  Casse, Goodman, Parsley, Pogrebenko, Schilizzi, \& Smythe}]{Whitney_2004}
Whitney, A.~R., Cappallo, R., Aldrich, W., {et~al.} 2004, Radio Science, 39,
  RS1007

\bibitem[{{Yuan} {et~al.}(2003){Yuan}, {Quataert}, \& {Narayan}}]{Yuan_2003}
{Yuan}, F., {Quataert}, E., \& {Narayan}, R. 2003, \apj, 598, 301

\bibitem[{{Zhu} {et~al.}(2019){Zhu}, {Johnson}, \& {Narayan}}]{Zhu_2018}
{Zhu}, Z., {Johnson}, M.~D., \& {Narayan}, R. 2019, \apj, 870, 6

\end{thebibliography}

\end{document}